\documentclass[prd,superscriptaddress,twocolumn,preprintnumbers,showpacs,nofootinbib]{revtex4-1}
\pdfoutput=1
\usepackage{graphicx,amssymb,amsmath}\usepackage{subfig}

\begin{document} \title{Longitudinal top polarisation measurement and anomalous $Wtb$ coupling}

\author{ArunPrasath V}\email{arunprasath@cts.iisc.ernet.in}
\author{Rohini M. Godbole} \email{rohini@cts.iisc.ernet.in}
\affiliation{Centre for High Energy Physics, Indian Institute of
Science, Bangalore 560 012, India}

\author{Saurabh D. Rindani} \email{saurabh@prl.res.in}
\affiliation{Physical Research Laboratory, Navarangpura, Ahmedabad, 380
009, India}

\begin{abstract}
Kinematical distributions of decay products of the top quark carry
information on the polarisation of the top as well as on any possible
new physics in the decay of the top quark. We construct observables in
the form of asymmetries in the kinematical distributions to probe their
effects. Charged-lepton angular distributions in the decay 
are insensitive to anomalous couplings to leading order. Hence these
can be a robust probe of top polarisation. 
However, these are difficult to measure in the case of
highly boosted top quarks as compared to energy distributions of decay products. 
These are then sensitive, in general, to both top polarisation and top
anomalous couplings. We compare various asymmetries for their sensitivities to the
longitudinal polarisation of the top quark as well as to possible new physics in
the $Wtb$ vertex, paying special attention to the case of highly boosted
top quarks. We perform a $\chi ^2$- analysis to determine the regions in
the longitudinal polarisation of the top quark and the couplings of the $Wtb$ vertex
constrained by different combinations of the asymmetries. Moreover, we find 
that use of observables sensitive to the longitudinal top polarisation can add to the 
sensitivity to which the $Wtb$ vertex can be probed.     
\end{abstract} 


\preprint{} 
\maketitle

\section{Introduction}\label{sec:int} 
The top quark is the heaviest of
all fundamental particles discovered so far in the Standard Model(SM).
Since the mass of the top quark ($m_t=173.5$
$\mathrm{GeV}/\mathrm{c}^2$) \cite{Beringer:1900zz} is very close to the electroweak
symmetry breaking (EWSB) scale, effects of any New Physics (NP)
associated with EWSB are likely to reveal themselves  in the properties
of the top quark. The LHC, during the course of its runs, is 
expected to determine several of the
properties of the top quark
\cite{Beneke:2000hk,*Han:2008xb,*Bernreuther:2008ju,*Schilling:2012dx,*BARBERIS:2013wba,*JABEEN:2013mva}. A comparison of
these with expectations from the SM will reveal NP, if present. 
In the search for NP, there are already
some results from the LHC, which include those 
on the top-quark polarisation and anomalous couplings in
the $Wtb$ vertex \cite{Aad:2013ksa,CMS-PAS-TOP-11-020}, which are
relevant to the discussion in this paper.  
 
 New Physics may appear in the production of the top quark or its decay
or both \cite{[{see for example}]Cheung:1995nt,*PhysRevD.54.4326,*PhysRevD.61.119901,*Godbole:2002qu,*Choudhury:2009wd,*Gupta:2009wu,*HIOKI:2011xx,*Choudhury:2012np,Antipin:2008zx,Bernreuther:2013aga}.
A model-independent way of probing NP in the top sector is provided by
the effective-theory formalism where all gauge-invariant higher-dimensional 
operators suppressed by powers of the corresponding scale of
NP are added to the SM Lagrangian
(\cite{Buchmuller:1985jz,*Grzadkowski:2003tf,*Grzadkowski:2010es,AguilarSaavedra:2008zc,Bach:2012fb,Zhang:2010dr}).
This description is valid at scales much lower than the NP scale.  A complete set of dimension-six  operators
relevant to top production and decay can be found
in \cite{Zhang:2010dr}. Higher order effects within the SM itself could induce structures that are not present at the tree level vertex (see for example, \cite{Fischer:2001gp,Czarnecki:1990kv,*Li:1990qf,GonzalezSprinberg:2011kx}).     

 The top quark, on account of its large mass, decays before it
hadronises, thereby transferring its spin information to the decay
products. The angular and energy distributions of the decay products carry information on the spin of the top quark
\cite{Jezabek:1988ja,Czarnecki:1990pe,Kane:1991bg,Bernreuther199253,*Erratum}. Kinematic distributions of the decay products of the top in the presence of anomalous couplings  have been studied, without assuming any model, in \cite{Jezabek:1994zv,Nelson:1997xd,Godbole:2006tq,Rindani:2000jg}. Effects of higher order QCD corrections on the distibutions are studied in \cite{Jezabek:1988ja,Czarnecki:1990pe,Jezabek:1988iv,Fischer:2001gp,Brandenburg:2002xr,Bernreuther:2003xj,Groote:2006kq,Hagiwara:2007sz,Kitadono:2012qk,Brucherseifer:2013iv,Bernreuther:2014dla}.

The polarisation of top quarks produced in a hadron collider like the
LHC depends upon the hard subprocesses that produce them. Since QCD which is mainly responsible for top-pair production in the SM is a vector interaction, there is no significant longitudinal polarisation of the top quarks pair produced in the SM: less than about a percent, after taking into account the one loop electro weak radiative corrections, in the so-called helicity basis\footnote{Helicity basis is defined as the basis in which the top spin quantisation axis is taken along the direction of motion of the top.} defined in $t\bar{t}$ centre of mass (c.m) frame of $pp$ collisions at the LHC \cite{Bernreuther:2006vg,*Bernreuther:2008md,Kuhn:2006vh,*PhysRevD.91.014020}.  Single-top production, occurring via electro-weak interactions at much lower rates, does give rise to polarised top quarks\cite{Mahlon:1999gz,Schwienhorst:2010je}. The value of polarisation after including NLO QCD corrections is $\sim$ 0.91 in the helicity basis defined in the c.m frame of the top quark and the spectator jet (the jet from the light quark that is scattered away along with the top) \cite{Schwienhorst:2010je}. Note that 
the top can be polarised along a direction that is perpendicular to the plane of production of the top (`transverse polarisation'). While in the SM longitudinal polarisation requires parity violating interactions, transverse polarisation is allowed even when the interactions are parity conserving as in the case of QCD. This however, is generated only at one loop level in QCD. In the case of top pair production, it does not exceed  $\sim 2\%$ at the parton level \cite{Dharmaratna:1989jr,*PhysRevD.45.124,Bernreuther:1995cx}. This value is further reduced at the LHC due to the dominance of the gluon fusion channel in $t\bar{t}$ production: $\sim 0.2\%$ at 7 TeV \cite{Baumgart:2013yra}. In this work,  we focus only on the longitudinal polarisation. Hence, in our paper the term top polarisation always refers to the longitudinal polarisation unless it is explicitly stated otherwise. 

 Since the standard model value of top polarisation in top pair production is small, an observation of substantial polarisation 
in top-pair production will strongly indicate NP. Any nontrivial chiral structure in the top coupling induced by the NP  can affect the polarisation of the produced top quarks
\cite{Nojiri:1994it,*Perelstein:2008zt,*Arai:2010ci,Gopalakrishna:2010xm,*Godbole:2010kr,Huitu:2010ad,*Rindani:2011pk,*Biswal:2012dr,*Belanger:2012tm,*Belanger:2013gha,*Taghavi:2013wy,*Aguilar-Saavedra:2014eqa,Bernreuther:2013aga}.Hence measurement of the polarisation of the top quark can provide information on the chiral structure of the couplings involved in NP contributions to top quark-production \cite{Nojiri:1994it,*Perelstein:2008zt,*Arai:2010ci,Gopalakrishna:2010xm,*Godbole:2010kr,Huitu:2010ad,*Rindani:2011pk,*Biswal:2012dr,*Belanger:2012tm,*Belanger:2013gha,*Taghavi:2013wy,Cao:2010nw,*Choudhury:2010cd,*Krohn:2011tw,Falkowski:2011zr,*Godbole:2011vw}.

New physics reflects itself in changes in total and differential cross
sections for top production. Detailed study of angular distributions of
the decay products of the top quark,  which are also affected by 
top-quark polarisation, provides a useful handle for discrimination between different NP models. Moreover, when NP couplings are small and the deviations of the total cross section from theoretical predictions in the SM can be small, the kinematic distributions and final-state polarisations being sensitive to the interference between the SM contribution and NP contribution can  lead to increased sensitivity.  
The top-quark polarisation can, in addition, give a handle on the chiral structure of the couplings in NP. 

A number of interesting scenarios for the production of top quarks occur
once extensions of the standard model are introduced. The most popular
extensions include supersymmetry, theories with extra dimensions and theories with extended gauge groups, all of which introduce new particles, which would contribute to top quark production in various ways: through an on shell production of resonances or via virtual effects. As said before, a nontrivial chiral structure of the top-quark couplings induced by the NP  will lead to a prediction for top-quark polarisation which depends on the values of the parameters of that particular extension of the SM being considered.

Some of the NP models predict new heavy resonances with masses at the
TeV scale~\cite{Randall:1999ee,*Agashe:2004rs,*Agashe:2006hk}. 
Such heavy resonances are produced effectively at rest in
the parton centre-of-mass (cm) frame. When these heavy resonances decay into top quarks, the resulting top quarks are highly boosted in the lab frame.
The decay products of these highly boosted top quarks
are collimated along the direction of motion of the parent top quark. In such a case observables based on the energy distributions of the decay products rather than their angular distributions are more suitable to probe the
polarisation of the top quark \cite{Shelton:2008nq}. 
For such  highly boosted tops methods based on jet substructure have been proposed to extract information on the polarisation of the top which then can be used to get information on the production mechanism of top quarks
\cite{Krohn:2009wm,*Bhattacherjee:2012ir,*Kitadono:2014hna}.  Recently a new method for measuring the polarization of top, when the top decays hadronically, has been proposed \cite{Tweedie:2014yda}. This method, involving a weighted average, in the top rest frame, of the directions of two light-quark jets that come from the decay $t\rightarrow bjj$, has been shown to perform better than methods based on other hadronic top spin analysers.

Data from the LHC has placed stringent lower bounds on the masses of
resonances \cite{Chatrchyan:2012cx,*Chatrchyan:2012oaa,*Aad:2012em,*Aad:2012raa,*Aad:2013nca}. If they do exist at higher masses, the observation in the invariant mass 
distribution would be difficult. On the other hand, NP
production amplitude of the resonance giving rise to a top pair could
have sizable interference with the SM amplitude. This could lead to
observable top polarization provided the NP couplings have a nontrivial
chiral structure. Top polarization can serve as a tool in testing these models ~\cite{Gopalakrishna:2010xm,*Godbole:2010kr}.

Another example of significant top polarisation is in stop decay
\begin{equation}
\tilde t_1 \to t \tilde \chi^0_i,
\end{equation}
in the minimal supersymmetric standard model (MSSM), where $\tilde t_1 $
is the lightest stop and $\tilde
\chi^0_i$, $i=1$,...,4 stand for the four neutralinos, which can be used to
study mixing in the sfermion sector as well as the neutralino-chargino
sector \cite{Perelstein:2008zt,Low:2013aza,Belanger:2013gha,Belanger:2012tm}. In $R$-parity violating MSSM, top quarks pair produced via a
$t$-channel exchange of a stau or a stop or a top quark produced in
association with a slepton, can have nonzero polarisation, whose
measurement can be used to constrain the $R$-parity violating couplings \cite{Hikasa:1999wy,Nie:2005fa,Arai:2010ci}.
There have been several NP explanations of the forward-backward
asymmetry of the
top quark observed at Tevatron (see, for example  \cite{Jung:2009jz,*Frampton:2009rk,*Shu:2009xf,*Cao:2010zb,*Dorsner:2009mq,*Cheung:2009ch,*Djouadi:2009nb,*Arhrib:2009hu,Bai:2011ed,*Gresham:2011pa,*Buckley:2011vc,*Bhattacherjee:2011nr,*AguilarSaavedra:2011vw,*AguilarSaavedra:2011ug,*Chivukula:2010fk,*Grinstein:2011yv,*Tavares:2011zg,*Jung:2011zv,*Ligeti:2011vt,Biswal:2012mr,*Allanach:2012tc,*Dupuis:2012is,*Aguilar-Saavedra:2014yea,Baumgart:2013yra}), and top polarisation can be useful in
discriminating among them \cite{Choudhury:2010cd,Krohn:2011tw,Cao:2011hr,Fajfer:2012si}. In fact the top polarisation transverse to the $t\bar{t}$ production plane also could be used to test whether the measured forward-backward asymmetry at the Tevatron is due to the effect of some NP in the top pair production \cite{Baumgart:2013yra}, even when the NP is difficult to be observed directly.

Since the top quark mainly decays through the channel $t\rightarrow Wb$
with a branching ratio of $\approx 100 \% $, any new physics which
appears through the $Wtb$ vertex can affect the measurement of
polarisation of top quarks which is determined by the production
process. 
In general, measures of top polarisation 
have a dependence both on the strength and the tensor structure of 
the $Wtb$ coupling  associated with top decay.   
 Measures of top polarisation which  depend only on the energy integrated angular distributions are insensitive to the anomalous part of the decay vertex $Wtb$
\cite{Grzadkowski:1999iq,*Grzadkowski:2001tq,*Grzadkowski:2002gt,Rindani:2000jg,Godbole:2006tq}.
Recently another measure of top-quark polarisation has been proposed  in \cite{AguilarSaavedra:2012xe} which factors out the the effect of any possible anomalous $Wtb$  vertex from the polarization of the top quark. The factor which contains the information about the $Wtb$ vertex of the top decay can be extracted in a model-independent way from the angular distributions of the top quark's decay products \cite{AguilarSaavedra:2006fy}. 
Since the anomalous $tbW$ coupling also affects the kinematic distributions of the decay products of the top quark, it too can
be probed by studying these and such probes have been constructed
\cite{Boos:1999dd,delAguila:2002nf,Hubaut:2005er,Najafabadi:2006um,AguilarSaavedra:2006fy,AguilarSaavedra:2008gt,Najafabadi:2008pb,AguilarSaavedra:2007rs,AguilarSaavedra:2010nx,Rindani:2011pk,Rindani:2011gt}. Probing anomalous $Wtb$ couplings at future colliders such as LHeC and ILC have also been considered by various authors\cite{Boos:1999ca,Dutta:2013mva,Lin:2001yq,Devetak:2010na}.

Since the NP can affect top polarisation as well as give
rise to anomalous decay vertex, it is of interest to explore how well one can study simultaneously both the  top polarisation and  the  anomalous $Wtb$ couplings and further see how probes of one
are influenced by the other. We present in this note some observations on construction of various observables as a measure of top polarisation and how one can simultaneously probe top polarization and the anomalous $Wtb$ coupling, when neither of the two is known a priori.

Studies of spin effects in top physics have  largely concentrated on
spin correlations in top pair production, as these are nonzero even in the SM and the  measurements are interesting, even if no NP effects exist. A comparison of experimental results with SM predictions can then be used to constrain the NP models. The results so obtained at the Tevatron and the LHC have so far shown consistency with the SM,
though errors are large. These correlations are best measured using
leptonic final states from both top and anti-top. It is
conceivable that a single polarisation measurement on either the top or
the anti-top which decays leptonically, allowing the other to decay
hadronically, could add to the accuracy.

Moreover, attempts to measure single-top polarisation at the Tevatron
and at the LHC have so far been made by reconstructing
the rest frame of the top quark. A method which does not require such full reconstruction of the top may be desirable. We have thus
concentrated on the measures of the polarization of a single top in the laboratory frame.

We construct various kinematic observables (asymmetries),
make a comparative study of their dependence on top polarisation and
anomalous $Wtb$ vertex, and examine the possibility of simultaneously
constraining the anomalous $Wtb$ vertex and the polarisation of the top
quark. 
Our observables do not always require full reconstruction of the top momentum. We do not look at any specific top production mechanism, but simply consider the top quark to be produced in the lab frame with
various momenta, paying special attention to highly boosted top quarks.

Our paper is divided into four sections. In section II we describe
the structure of $Wtb$ vertex and constraints on various anomalous
couplings. In section III we make a comparative study of the
sensitivities of  different asymmetries to the polarisation of the top
and anomalous $Wtb$ couplings. In section IV we use these asymmetries
to constrain simultaneously the polarisation of the top quark and the
$Wtb$ vertex. In section V we give a summary and conclusions.
\footnote{In this work all kinematic quantities in  the rest  frame of
the top quark are denoted by a subscript `0'. All kinematic quantities
which do not have subscript `0' are in the laboratory frame (lab frame),
unless stated otherwise. We have assumed that the top quark has spin along the direction of motion of the top in the lab frame.  The lab frame polarisation is obtained from the rest frame one by a boost along the direction of motion of the top quark. We use the word lepton to denote the charged anti-lepton $\bar{\ell}$ from $t\rightarrow b\bar{\ell}\nu$.}        

\section{The structure of $Wtb$ vertex}\label{sec:wtb} 

The $Wtb$ vertex
in the SM has a $V-A$ structure. Depending upon the NP the structure of
$Wtb$ vertex can be modified from the $V-A$ structure \cite{[{see, for
example }]Bernreuther:2008us}. We follow a model-independent approach by
writing down the most general $Wtb$ vertex \cite{AguilarSaavedra:2006fy}: 
\begin{align}
\Gamma ^{\mu}&=-i\frac{g}{\sqrt{2}}\big[\gamma
^{\mu}(f_{1L}P_L+f_{1R}P_R)\notag\\ &-\frac{i}{M_W}\sigma
^{\mu\nu}(p_t-p_b)_{\nu}(f_{2L}P_L+f_{2R}P_R)\big]\label{eq:1} \end{align} 
where $g$
is the $SU(2)_L$ gauge coupling constant, $p_t,p_b$ are the four-momenta
of the top and the bottom quarks respectively, and $P_L,P_R$ are the left and
right chiral projectors.  In the SM, $f_{1L}=1$ and
$f_{1R}=f_{2L}=f_{2R}=0$ at tree level. One loop QCD contributions to the $Wtb$ vertex have been computed  within the SM, in\cite{Czarnecki:1990kv,*Li:1990qf,*Brandenburg:2002xr}
. Electroweak corrections also affect the $Wtb$ vertex and in turn affect the couplings $f_{1L,R}$ and $f_{2L,R}$. One loop EW contributions to the tensor couplings amount to about $10\%$ of the one loop QCD contributions \cite{GonzalezSprinberg:2011kx}. After including the EW contributions the tensor couplings at one loop level in the SM are as follows: $f_{2L}=-(1.21+0.01i)\times 10^{-3}$ and $f_{2R}=-(7.17+1.23i)\times 10^{-3}$\cite{GonzalezSprinberg:2011kx}.  

We take the CKM matrix element $V_{tb}\approx 1$. Similarly, the vertex $\bar{t}W\bar{b}$ with anomalous couplings is given by 
\begin{align}
\Gamma ^{\mu}&=-i\frac{g}{\sqrt{2}}\big[\gamma
^{\mu}(\bar{f}_{1L}P_L+\bar{f}_{1R}P_R)\notag\\ &-\frac{i}{M_W}\sigma
^{\mu\nu}(p_t-p_b)_{\nu}(\bar{f}_{2L}P_L+\bar{f}_{2R}P_R)\big]\label{eq:1a} \end{align} 

When CP is conserved, $f_{1L}=\bar{f}_{1L}$, $f_{1R}=\bar{f}_{1R}$, $f_{2L}=\bar{f}_{2R}$ and $f_{2R}=\bar{f}_{2L}$ ~\cite{Rindani:2011gt}. Direct searches of NP in top decay at the Tevatron
give constraints on the coefficients: $|f_{1R}|^2<0.30$,
$|f_{2L}|^2<0.05$, $|f_{2R}|^2<0.12$ at 95 \% C.L assuming $f_{1L}=1$
\cite{Abazov:2012uga}. Indirect constraints from the measurement of the
branching ratio of $b\rightarrow s \gamma $ are stronger for $f_{1R}$,
$f_{2L}$ and weaker for $f_{2R}$:$-0.15\leq \operatorname{Re}(f_{2R})\leq 0.57$,
$-0.0007\leq f_{1R}\leq 0.0025$, $-0.0013\leq f_{2L}\leq 0.0004$
\cite{Grzadkowski:2008mf} at 95\% C.L. Direct search constraints are
given also by the LHC: for $f_{1L}=1$, $f_{1R}=f_{2L}=0$ the CMS
collaboration \cite{CMS-PAS-TOP-11-020} obtained as a best fit value
of the tensor part of the $Wtb$ coupling which in our notation reads as $f_{2R}=0.070\pm 0.053\,(\mathrm{stat})^{+0.081}_{-0.073}\,(\mathrm{syst})$ in a fit to
measured $W$ helicity fractions proposed by \cite{Kane:1991bg}. A combination of  data from the Tevatron, and the LHC on observables like the $t$-channel single top cross section, and the $W$ helicity fractions, gives, in our sign conventions, $-0.11\leq\operatorname{Re}(f_{2R})\leq 0.13$ and $-0.31\leq\operatorname{Im}(f_{2R})\leq 0.31$ respectively at 95$\%$ C.L\cite{Bernardo:2014vha}. When CP is conserved, the constraints on the anomalous couplings $\bar{f}_{1L}$, $\bar{f}_{1R}$, $\bar{f}_{2L}$ and $\bar{f}_{2R}$ are the same as those for $f_{1L}$, $f_{1R}$, $f_{2R}$ and $f_{2L}$ respectively.

In the analytical expressions for different kinematic distributions (see below) we have assumed that the anomalous couplings $f_{1L}$ and $\bar{f}_{1L}$ are real valued while all other anomalous couplings are complex valued. However, in view of the strong constraints on the anomalous couplings, in our numerical work and in analyical results on  $Wtb$ vertex, we set  $f_{1R}=f_{2L}=0$. In numerical works, we also set $f_{1L}=1$ and take $f_{2R}$ as a real valued quantity varying in the range $-0.2$ to $+0.2$. 

\section{Kinematic distributions  of the decay products of the top}\label{sec:1}

Recall that a measurement of the polarisation of the top quark can only be done through the kinematic distributions of its decay products and  these would also be affected by the nature of the $Wtb$ coupling.

We begin by looking at the details of the three-body decay of the top quark. 
The top quark decays into a $b$ quark and a $W$
boson which in turn decays into a charged anti-lepton ($\bar{\ell}$) and
a neutrino $\nu _{\ell}$. We assume that all the particles in the decay
chain $t\rightarrow bW\rightarrow b\bar{\ell}\nu _{\ell}$ are on-shell
(including the intermediate $W$ boson). The angular distributions of the
decay products are correlated to the polarisation $P$ of the top quark. In the SM, in the rest frame of the top quark, the energy integrated  distribution is given by 
\begin{equation}\label{eq:spin}
\frac{1}{\Gamma}\frac{d\Gamma}{d\cos\theta _X}=\frac{1}{2}(1+P\alpha
_X\cos\theta _X) 
\end{equation} 
where $X=b,\ell ,W,\nu _{\ell}$, the
quantity $\alpha _X$ is called the spin-analysing power of the particle
$X$ and $\theta _X$ is the angle between the direction of the momentum
of the particle $X$ and the top quark spin axis in the rest frame of the
top quark. The spin-analyzing powers of the $b$ quark, lepton and the
neutrino in the SM at tree level are 
\begin{align*} 
\alpha_b &=-\left(\frac{\xi -2}{\xi
+2}\right)\\ \alpha _{\ell} &=1\\ \alpha _{\nu}&= 1-\frac{12\xi(\xi
-1-\log\xi)}{(\xi -1)^2(\xi +2)}\\ 
\end{align*} 
respectively, where
$\xi=\frac{m_t^2}{m_W^2}$ \cite{Jezabek:1994qs}. Higher order QCD effects on the spin analysing power of the lepton $\ell$ are at the per-mille level \cite{Czarnecki:1990pe}.  But  in the case of hadronic decay of the top, the QCD corrections to the spin analysing powers of quarks from the top decay can be upto about  $4\%$ \cite{Brandenburg:2002xr,Bernreuther:2014dla}.

\subsection{Kinematics of the top
decay}\label{ssec:kin} Before proceeding to the description of
asymmetries, we give a brief description of the kinematics of the top
quark decay: $t\rightarrow Wb\rightarrow b\bar{\ell}\nu$ . 

  The conservation of energy and momentum and the on-shell condition of
$W$ give the following equations: 
\begin{equation}\label{eq:3a1}
p_t=p_b+p_{\ell}+p_{\nu} 
\end{equation} 
and 
\begin{equation}\label{eq:3a2}
m_W^2=(p_{\ell}+p_{\nu})^2 
\end{equation} 
where
$p_t,p_b,p_{\ell},p_{\nu}$ are the four-momenta of the particles.
Solving these equations along with the on-shell condition of the
particles  fixes all but four variables in the rest frame of the top
quark: energy of the lepton $E_{\ell,0}$, the polar and azimuthal angles
of the lepton $\theta _{\ell,0},\phi _{\ell,0}$ and the azimuthal angle
($\alpha _0$) of the $b$ quark with respect to a coordinate system where
the $z$ azis is along the direction of the lepton momentum. The variables fixed by eqs.~\ref{eq:3a1} and ~\ref{eq:3a2} and the on-shell conditions are, the energy of the $b$ quark, $E_{b,0}=(m_t^2-m_W^2)/2m_t$ and the cosine of the
angle between the $b$-quark momentum and the lepton momentum,
$\cos\zeta=(2-x_{\ell,0}(\xi +1))/(x_{\ell,0}(\xi -1))$ where,
$\xi=m_t^2/m_W^2$. The energy $E_{\ell,0}$ of the lepton is constrained to
vary between $m_W^2/2m_t$ and $m_t/2$.  \subsection{Definition of
asymmetry}\label{ssec:asymdef}  Asymmetry in a kinematic variable $X$ is
defined (in a given frame) as 
\begin{equation}\label{eq:2}
A_X=\frac{\int_{X_{min}}^{X_{c}}\frac{d\Gamma}{dX}dX-\int_{X_{c}}^{X_{max}}\frac{d\Gamma}{dX}dX}{\int_{X_{min}}^{X_{c}}\frac{d\Gamma}{dX}dX+\int_{X_{c}}^{X_{max}}\frac{d\Gamma}{dX}dX}
\end{equation} 
where $\frac{d\Gamma}{dX}$ is the
differential partial decay width of the top quark in the variable $X$
and $X_c$ is a value of $X$ between
$[X_{\mathrm{min}},X_{\mathrm{max}}]$ chosen as a reference point about
which the asymmetry is evaluated. In this work, we chose reference points on the basis of the following considerations:
\begin{enumerate}
\item The reference point should be such that the evaluated asymmetry is sensitive to both of the parameters $P$ and $f_{2R}$ throughout their allowed range of values. In other words, the asymmetries thus obtained should be able to distinguish between different values of the parameters.
\item The choice of the reference point should allow for a comparison of cases of different values of the parameters.
\end{enumerate} 

  The asymmetries vary in their sensitivity to the polarisation of the top
quark and the anomalous coupling $f_{2R}$, and in their 
usefulness in a particular
kinematic regime of top decay. We describe four such asymmetries in
this section. We can classify them into broadly two categories: angular
asymmetries and energy-based asymmetries. Examples of the former include
$A_{\theta _{\ell}}$ and of the latter include $A_{x_{\ell}}$, $A_{u}$, and 
$A_z$. When the top quarks are highly boosted, the decay products of the
top are highly collimated. In this case measurement of  angular
distribution of visible decay products is possible but
difficult \cite{ATL-PHYS-PUB-2010-008}. Hence energy-based asymmetries
are used to probe the top-quark polarisation \cite{Shelton:2008nq} and
$Wtb$ vertex. 
 
\subsection{The $\theta _{\ell}$ asymmetry  ($A_{\theta
_{\ell}}$)}\label{ssec:1} The asymmetry is defined in terms of
$\cos\theta _{\ell}$ where $\theta _{\ell}$ is the angle between the
momentum of the lepton from the $W$ decay and the top quark direction of
motion. The $\cos\theta _{\ell}$ distribution is sensitive to the
polarisation of the top quark ($P$). In the rest frame of the top quark
the distribution is given  in the SM by
\begin{equation}\label{eq:apthlcm}
\frac{1}{\Gamma}\frac{d\Gamma}{d\cos\theta
_{\ell,0}}=\frac{1}{2}(1+P\cos\theta _{\ell,0}). 
\end{equation} 
This
expression is valid even when the anomalous coupling $f_{2R}$ is non
zero provided it is small
\cite{Grzadkowski:1999iq,*Grzadkowski:2001tq,*Grzadkowski:2002gt,Rindani:2000jg,Godbole:2006tq}.
In the lab frame  the $\cos\theta _{\ell}$ distribution becomes \cite{Rindani:2011pk}
\begin{align}\label{eq:3}
\frac{1}{\Gamma}\frac{d\Gamma}{dt}&=\frac{(1-\beta ^2)}{2(1-\beta
t)^3X}[ (\xi -1)^2(f_{1L}^2(\xi+2)\\ \nonumber
&+6\sqrt{\xi}f_{1L}\operatorname{Re}(f_{2R}))
\{P(t-\beta)+(1-\beta t)\}\\ \nonumber
&+12P\xi(t-\beta)|f_{2R}|^2(-\xi+1+\xi\log\xi)\\ \nonumber 
&+|f_{2R}|^2(\xi-1)^2(2\xi+1)\{(1-\beta t)-P(t-\beta)\}
] 
\end{align} 
where $t=\cos\theta _{\ell}$ is
the cosine of the angle between the top quark direction of motion and
the lepton momentum in the lab frame. In eq. \ref{eq:3}, the factor $X$ 
in the denominator of the right-hand side
is given by
\begin{equation}\label{eq:4} 
X=(\xi -1)^2[(\xi
+2)f_{1L}^2+6\sqrt{\xi}f_{1L}\operatorname{Re}(f_{2R})+(2\xi +1)|f_{2R}|^2] 
\end{equation}
and $\beta$ is the boost required to go from the lab frame to the top-quark rest frame. When
$|f_{2R}|\ll 1$, keeping only terms which are first order in $f_{2R}$,
we get 
\begin{equation} 
\frac{1}{\Gamma}\frac{d\Gamma}{dt}\sim
\frac{(1-\beta ^2)}{2(1-\beta t)^3}((1-\beta P)+(P-\beta)t)
\end{equation} 
which is completely independent of the anomalous coupling
$f_{2R}$.  In other words, the energy-integrated distribution $1/\Gamma
d\Gamma/d\cos\theta _{\ell}$ is only very very weakly dependent on the
anomalous coupling $f_{2R}$.

 Therefore the lepton angular asymmetry ($A_{\theta _{\ell}}$) serves as
a useful measure of polarisation of the top quark irrespective of NP
effects at the decay vertex when they are small \cite{Godbole:2006tq}.

 In the SM, the asymmetry about the point $\cos\theta _{\ell}=0$ is
given in the lab frame, using eq. ~\ref{eq:2}, by
\begin{equation}\label{eq:6} 
A_{\theta _{\ell}}=\frac{1}{2}(-2\beta
+P(-1+\beta ^2)) 
\end{equation} 
From this equation one can easily
observe that  the sensitivity of $A_{\theta _{\ell}}$ to the
polarisation of the top quarks decreases with increasing boost. This can
be understood  as follows: In the rest frame of the top quark, the
lepton is preferentially emitted either in the forward direction or the
backward direction depending upon the sign of the polarisation of the top quark
($P$) (eq.~\ref{eq:apthlcm}). But in the lab frame, at large
values of boost, the lepton emission is strongly suppressed except in
the direction $\cos\theta _{\ell}=1$ due to kinematics which appears
through the factor $(1-\beta ^2)/(1-\beta t)^3$ in eq.~\ref{eq:3} for
all values of polarisation $P$. This means that the lepton angular
distribution loses its sensitivity to $P$ at large boosts as shown in eq.
~\ref{eq:6}.  \subsection{The $x_{\ell}$ asymmetry
($A_{x_{\ell}}$)}\label{ssec:2} 
\begin{figure}
\includegraphics[keepaspectratio=true,scale=0.4]{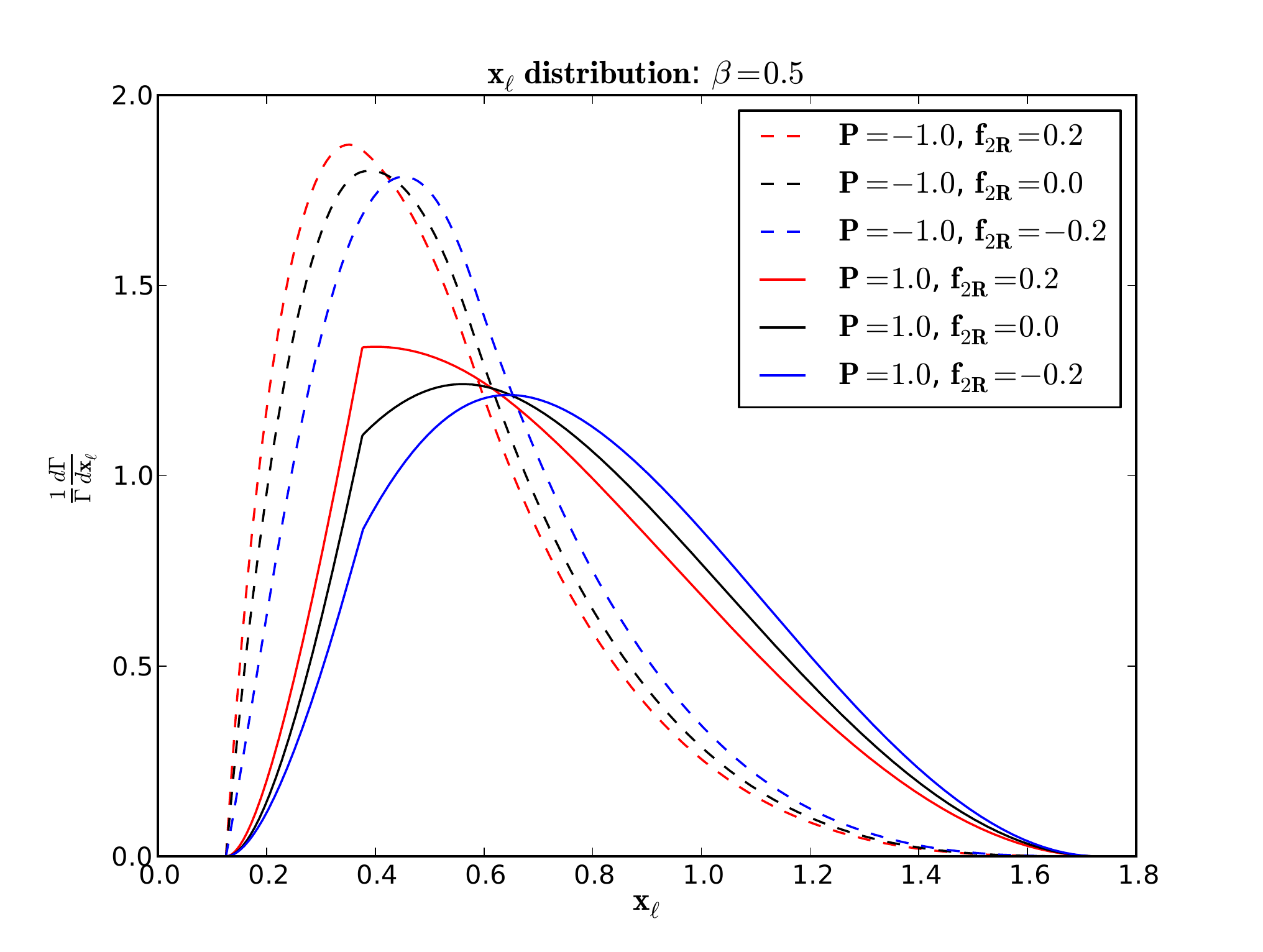}
\caption{The $x_{\ell}$ distribution in the lab frame for different
values  of the polarisation of the top quark ($P$) and the anomalous
coupling ($f_{2R}$).} \label{fg:el} 
\end{figure}

The variable $x_{\ell}$ is defined as
$x_{\ell}=2E_{\ell}/m_t$ where $E_{\ell}$ is the energy of the lepton
from the top decay in a given frame. Unlike the $\theta _{\ell,0}$
distribution, the $x_{\ell,0}$ distribution is not insensitive to
$f_{2R}$ in the top quark rest frame.  The analytical expression for the
distribution $(1/\Gamma) d\Gamma/dx_{\ell,0}$, in the top quark rest
frame, is given by 
\begin{align}\label{eq:7}
\frac{1}{\Gamma}\frac{d\Gamma}{dx_{\ell,0}}&=6\xi
^2\frac{1}{X}(1-x_{\ell,0})\\ \nonumber &[f_{1L}^2\xi
x_{\ell,0}+f_{1L}\operatorname{Re}(f_{2R})\sqrt{\xi}+|f_{2R}|^2(\xi +1-x_{\ell,0}\xi)]
\end{align} 
where $X$ is given in eq.~\ref{eq:4}. In the case when
$|f_{2R}|\ll 1$ the distribution is not independent of $f_{2R}$ as the
factors that are linear in $f_{2R}$ do not cancel each other from the
denominator and the numerator of eq.~\ref{eq:7}. The distribution
$1/\Gamma d\Gamma/dx_{\ell}$ is plotted in fig.~\ref{fg:el} for different
values of the top polarisation $P$ and the anomalous coupling $f_{2R}$.
The location of the  peak of the distribution for a given top
polarisation $P$ depends upon the anomalous coupling $f_{2R}$ as can be
seen from the figure. The sharp edges in the distribution for $P=1$ appear due to the interplay of the polarisation of the top and the kinematics of the top decay.
 
It would be convenient if the value of $x_{\ell}$ at the position of the 
 peak is chosen as the reference point to evaluate the asymmetry $A_{x_{\ell}}$. 
 In view of the fact that this point varies with $P$ as well as $f_{2R}$, for
 uniformity we take the value of $x_{\ell}$ corresponding to the peak of the
distribution for $P = -1$ and $f_{2R} =0$ as a reference point for all values
 of $P$ and $f_{2R}$. This choice is consistent with our method of choosing the reference points as  given in Sec. ~\ref{ssec:asymdef}.

 The above equation also shows that the $x_{\ell,0}$ distribution is
independent of  $P$ in the rest frame
of the top quark. Therefore the asymmetry  $A_{x_{\ell,0}}$ has no
sensitivity to the polarisation of the top quark ($P$) in the top quark rest
frame. But under a Lorentz transformation along the top quark direction
of motion which takes the top quark rest frame to the lab frame, the
energy of the lepton in the lab frame gets related to both the energy
and the polar angle $\theta _{\ell,0}$ of the lepton measured in the top
quark rest frame: 
\begin{equation*} 
E_{\ell}=\frac{1}{\sqrt{1-\beta ^2}}
E_{\ell,0}(1+\beta\cos\theta _{\ell,0}) 
\end{equation*} 
Since the
distribution in $\theta _{\ell,0}$ is correlated to the top quark
polarisation ($P$) (see eq. ~\ref{eq:apthlcm}) , the distribution in
$E_{\ell}$ (or $x_{\ell}$) becomes dependent on 
$P$. Hence the asymmetry $A_{x_{\ell}}$ for $\beta\neq 0$ depends on the
polarisation of the top quark ($P$). 

A variable similar to $x_{\ell}$ has been proposed in
\cite{Berger:2012an}. It is defined as $x_B=2E_{\ell}/E_t$ and it is
related to $x_{\ell}$ by a boost dependent factor: $x_B=\sqrt{1-\beta
^2}x_{\ell}/2$. However, the asymmetry constructed out of $x_B$-
distribution is the same as the asymmetry $A_{x_{\ell}}$ at any given
value of $\beta$.  \subsection{The $u$ Asymmetry ($A_u$)}\label{ssec:3}
The variable $u$ is defined as $u=E_{\ell}/(E_{\ell}+E_b)$ where
$E_{\ell}$ and $E_b$ are the energies in the lab frame 
carried by the lepton and the
$b$ quark respectively \cite{Shelton:2008nq}. The
variable $u$ can be written as 
\begin{equation}\label{eq:9} 
u=\frac{\xi
x_{\ell,0}(1+\beta\cos\theta _{\ell,0})}{\xi
x_{\ell,0}(1+\beta\cos\theta _{\ell,0})+(\xi -1)(1+\beta\cos\theta
_{b,0})} 
\end{equation} 
where $x_{\ell,0}=2E_{\ell,0}/m_t$ ,$E_{\ell,0}$
and $\theta _{\ell,0}$ are the energy and the angle between the top
quark direction of motion and the momentum of the lepton measured in the
top quark rest frame respectively. $\cos\theta _{b,0}$ is given by
\begin{equation}\label{eq:10} 
\cos\theta _{b,0}=-\sin\theta
_{\ell,0}\sin\zeta\cos\alpha _0 +\cos\theta _{\ell,0}\cos\zeta
\end{equation} 
with $\cos\zeta =\frac{2-x_{\ell,0}(\xi
+1)}{x_{\ell,0}(\xi -1)}$ and $0\leq\alpha _0\leq 2\pi$,$(1/\xi)\leq
x_{\ell,0}\leq 1$. $u$ varies in the range $(0,1)$. The
$u$ distribution is given by 
\begin{multline}\label{eq:11}
\frac{1}{\Gamma}\frac{d\Gamma}{du} = \int _0^{\pi}\int
_{1/\xi}^{1}dx_{\ell,0}d\theta _{\ell,0}\sum _{\alpha
_{0,i}}\frac{1}{J(\alpha _{0,i})}\left(\frac{3\xi
^2}{2\pi}\right)\frac{1}{X}\sin\theta _{\ell,0}\\
\hspace{-3cm}\times(1-x_{\ell,0})[f_{1L}^2\xi x_{\ell,0}(1+P\cos\theta _{\ell,0})\\ +f_{1L}\operatorname{Re}(f_{2R})\sqrt{\xi}[Px_{\ell,0}(\cos\theta _{b,0}(\xi
-1)+\cos\theta _{\ell,0}(\xi +1))+2]\\\hspace{0.1cm} +\hspace{0.2cm}|f_{2R}|^2(\cos\theta
_{b,0}P(\xi -1)+\cos\theta _{\ell,0}P\xi x_{\ell,0}+\xi +1
-x_{\ell,0}\xi)] 
\end{multline} 
where $X$ is as defined in eq.~\ref{eq:4}, and
$\alpha _{0,i}$,$(i=1,2)$ are the roots of the equation
$u=u(x_{\ell,0},\theta _{\ell,0},\alpha _0)$. 

Since $u$ is invariant under the
transformation $\alpha _0 \rightarrow 2\pi -\alpha _0$, we have two
solutions: $\alpha _{0,1}$ and $2\pi -\alpha _{0,1}$ with $0\leq \alpha
_{0,1}\leq \pi$. The function $J(\alpha _{0,i})$ is given by
\begin{equation}\label{eq:12} 
J(\alpha _{0,i})=-\frac{u^2(\xi
-1)\beta\sin\theta _{\ell,0}\sin\zeta\sin\alpha
_{0,i}}{x_{\ell,0}\xi(1+\beta\cos\theta _{\ell,0})} 
\end{equation} 
where
\begin{align}\label{eq:13} 
\cos\alpha _{0,i}&=\frac{(1+\beta\cos\theta
_{\ell,0}\cos\zeta)}{\beta\sin\theta _{\ell,0}\sin\zeta}\notag\\
&-\frac{(1/u-1)\xi x_{\ell,0}(1+\beta\cos\theta _{\ell,0})}{(\xi
-1)\beta\sin\theta _{\ell,0}\sin\zeta} 
\end{align} 
This equation
determines $\alpha  _{0,i}$ which can be used to get the value of the
distribution at $u$. The asymmetry $A_{u}$ is calculated the point $u=
u_c = 0.5$ using 
eq. ~\ref{eq:2}. 

We note that the $u$-distribution is independent of $\operatorname{Im}(f_{2R})$ to linear order: the integrand of eq.~\ref{eq:11} actually has an additional term  that is proportional to $P(1-x_{\ell})x_{\ell}\sin\alpha\sin\theta _{\ell}\sin\zeta$. Since the $u$-distribution is obtained after summing over two values of $\alpha$ i.e $\alpha _{0,1}$ and $2\pi -\alpha _{0,1}$, this additional term does not give any contribution. 

\begin{figure}
\includegraphics[keepaspectratio=true,scale=0.4]{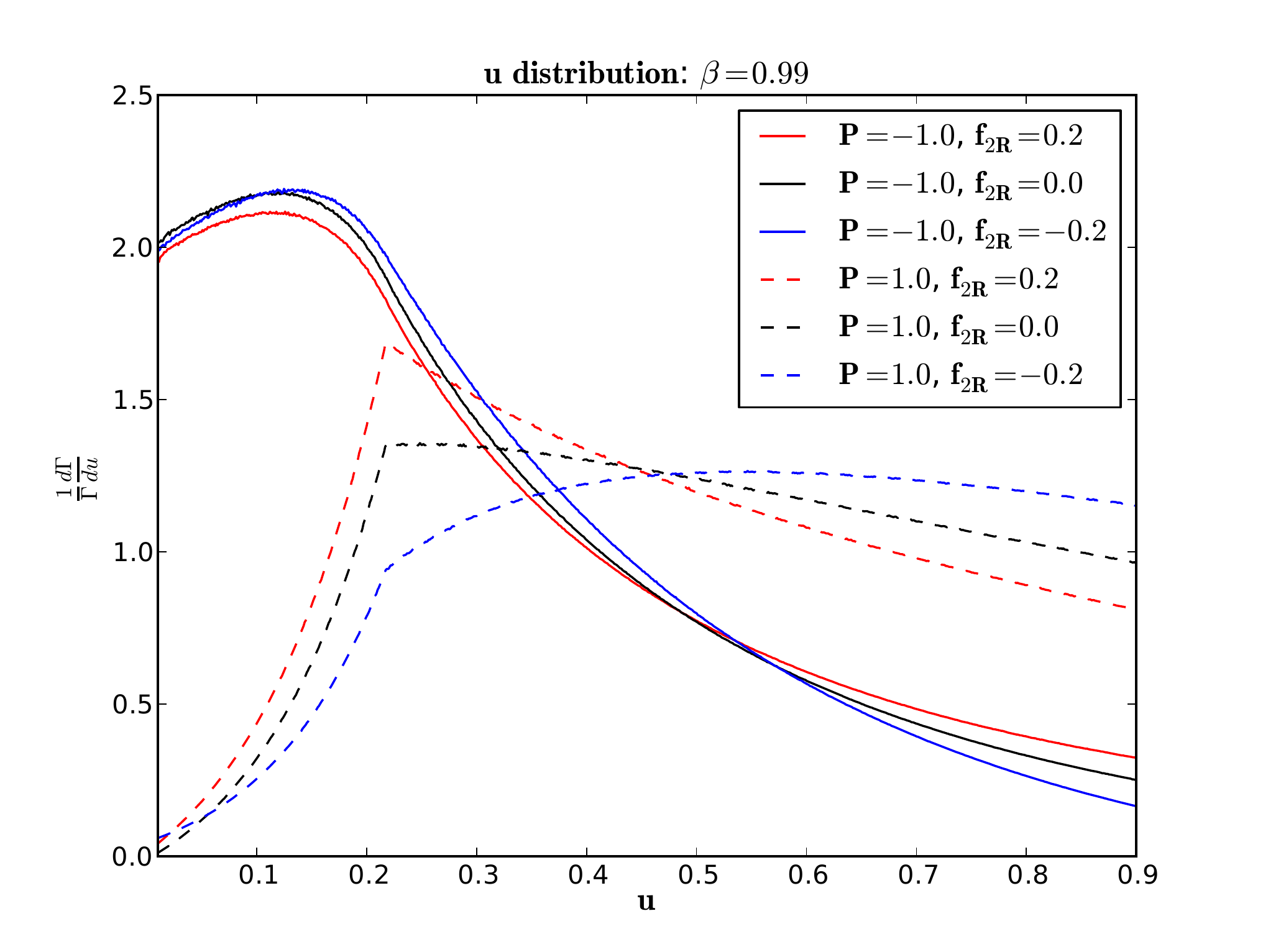}
\caption{The $u$ distribution as a function of the polarisation of the
top quark ($P$) and the anomalous coupling $f_{2R}$.} \label{fg:u}
\end{figure}

The $u$ distribution for
different value of $P$ and $f_{2R}$ is given in fig.~\ref{fg:u}. Similar to the case of 
$x_{\ell}$ distribution, the $u$ distribution has sharp edges due to the same reasons given in sec.~\ref{ssec:2}.  

\subsection{The $z$ asymmetry ($A_z$)}\label{ssec:4} 
\begin{figure*}[t!]
\includegraphics[keepaspectratio=true,scale=0.35]{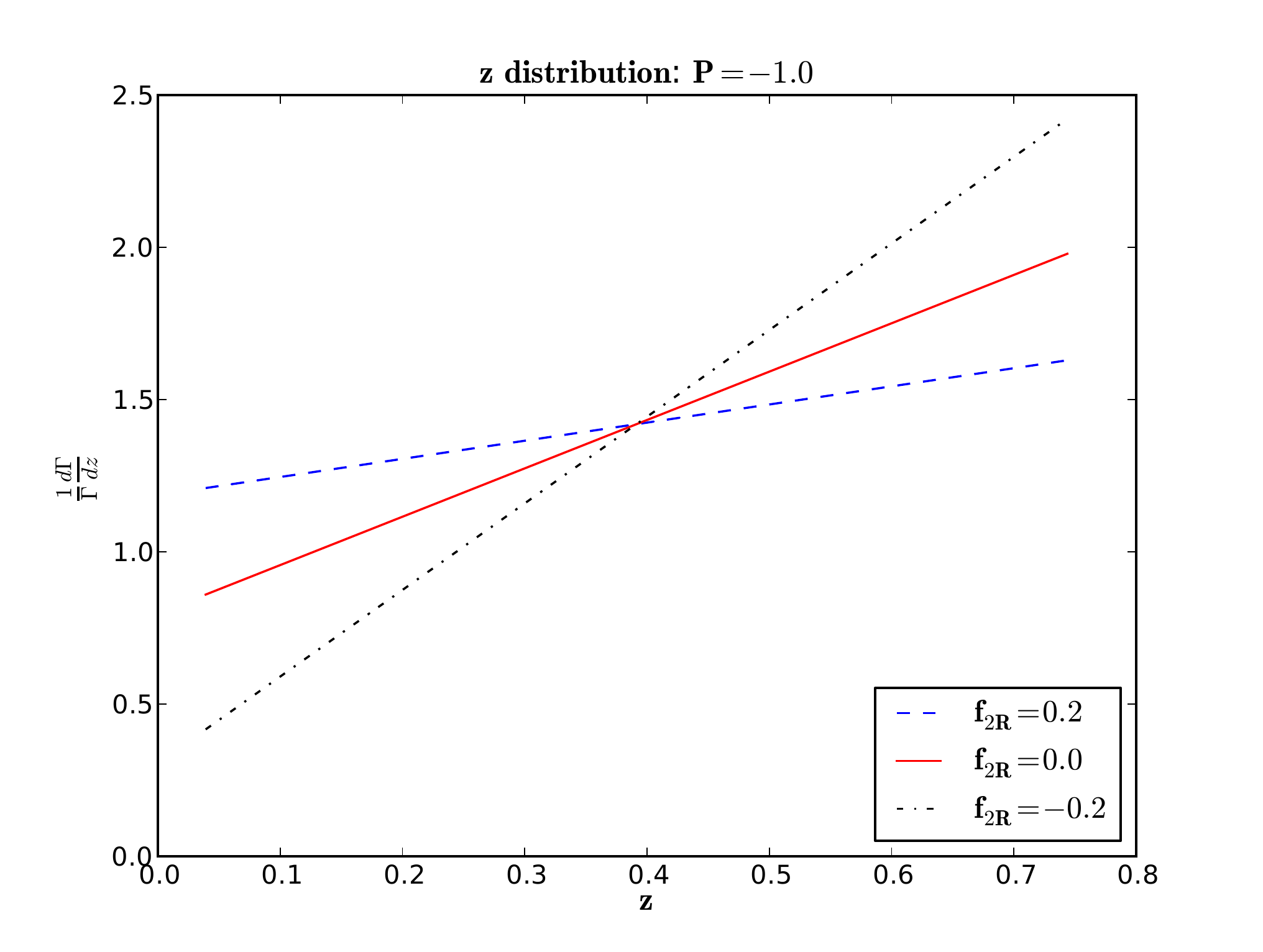}
\includegraphics[keepaspectratio=true,scale=0.35]{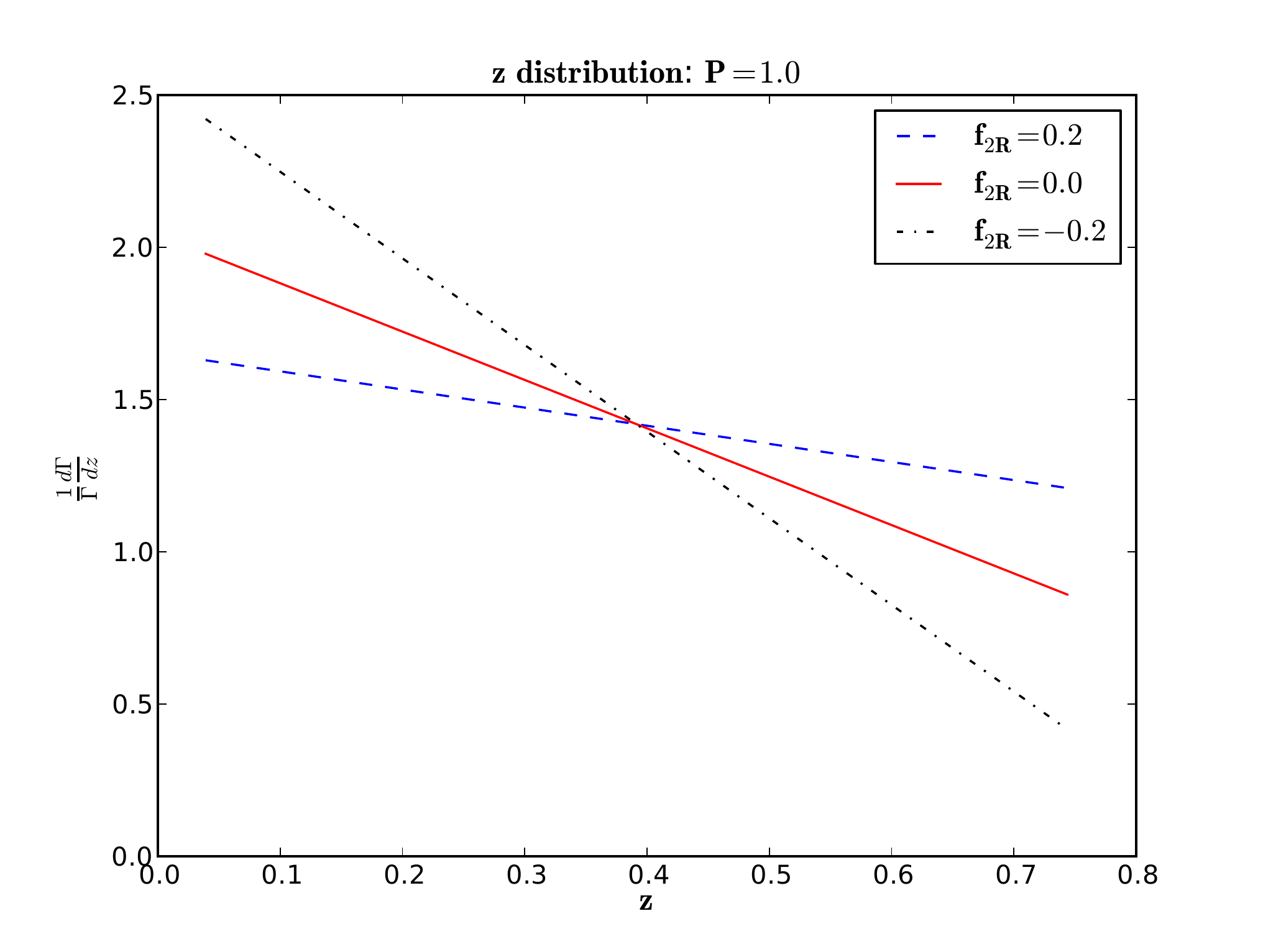}
\caption{The $z$ distribution $1/\Gamma d\Gamma/dz$ is plotted as a
function of $z$ for different values of $f_{2R}$. The left(right) figure
correspond to the polarisation of the top quark $P=-1.0$($P=1.0$). The
different lines show the $z$ distribution for different values of
$f_{2R}$. The solid, dashed and dash-dotted lines correpond to
$f_{2R}=0.0,0.2,-0.2$ respectively.} \label{fg:zdist} 
\end{figure*}

The variable $z$ is
defined as $z=E_{b}/E_t$ where $E_{b}$ and $E_{t}$  are the energies
in the lab frame 
carried by the $b$ and $t$ quarks respectively
\cite{Shelton:2008nq}. The variable $z$ can be related to the variables
defined in the rest frame of the top quark: 
\begin{equation}
z=\frac{(\xi -1)}{2\xi}(1+\beta\cos\theta _{b,0}) \label{eq:z-b}
\end{equation} 
where $\cos\theta _{b,0}$ is as defined in eq. ~\ref{eq:10}. 
Since $\cos\theta _{b,0}$ varies in the range
$[-1.0,1.0]$, the variable $z$ varies in the range $\left[(1-\beta)(\xi
-1)/2\xi,(1+\beta)(\xi -1)/2\xi\right]$. The $z$ distribution is given
by 
\begin{align}\label{eq:14} 
\frac{1}{\Gamma}\frac{d\Gamma}{dz}
&=\frac{\xi}{\beta ^2}\frac{1}{X}[\beta(\xi -1)\{f_{1L}^2(\xi +2)\notag\\
 &+6\sqrt{\xi}f_{1L}\operatorname{Re}(f_{2R})+|f_{2R}|^2(2\xi +1)\}\notag\\
&+P(-\xi +1+2z\xi )\{-f_{1L}^2(\xi -2)\notag\\
&+2\sqrt{\xi}f_{1L}\operatorname{Re}(f_{2R})+|f_{2R}|^2(2\xi -1)\}].
\end{align} 
The $z$ distribution is plotted in fig.~\ref{fg:zdist} for different values
of the top polarisation $P$ and the anomalous coupling $f_{2R}$. One can
easily observe that the effect of the anomalous coupling $f_{2R}$ is to
change the slope of the $z$ distribution which will be explained below.

 Since the distribution is linear in $z$, an analytical expression for
the asymmetry can be easily found. We take as the reference point $z_c$,
the value of $z$ which corresponds to $\cos\theta _{b,0}=0$ i.e
$z_c=(\xi -1)/2\xi$. To simplify the notation let us define two functions
of $f_{2R}$: $U=f_{1L}^2(\xi +2)+6\sqrt{\xi}f_{1L}\operatorname{Re}(f_{2R})+|f_{2R}|^2(2\xi
+1)$ and $V=-f_{1L}^2(\xi -2)+2\sqrt{\xi}f_{1L}\operatorname{Re}(f_{2R})+|f_{2R}|^2(2\xi
-1)$. Then the distribution in $z$ can be rewritten in terms of
$t_{b,0}=\cos\theta _{b,0}$ (see ~\cite{Rindani:2011pk,AguilarSaavedra:2010nx}): 
\begin{equation}
\frac{1}{\Gamma}\frac{d\Gamma}{dz}=\frac{1}{X}\left(\frac{2\xi(\xi
-1)}{\beta}\right)\left[\frac{U}{2}+\frac{PV}{2}t_{b,0}\right]
\label{eq:z-k} 
\end{equation} 
Now changing the variable from $z$ to
$t_{b,0}$ we get the limits of the integration in eq.~\ref{eq:2} as
$t_{b,0,\mathrm{min}}=-1$ and $t_{b,0,\mathrm{max}}=1$. Only the term
linear in $t_{b,0}$ survives in the numerator and the expression for
$A_z$ is given by 
\begin{equation} 
A_z=-P\frac{(\xi -1)^2V}{2X}
\end{equation} 
Therefore the asymmetry $A_z$ is directly proportional to
the top-quark polarisation $P$ and is independent of the boost factor
$\beta$ as $V$ and $X$ are independent of both $P$ and $\beta$. Moreover
the $z$ distribution can be directly related to the angular distribution
of the $b$ quark in the top rest frame due to eq.~\ref{eq:z-b}. In fact,
substituting the relation eq.~\ref{eq:z-b} in eq.~\ref{eq:z-k}, we
get, 
\begin{equation} 
\frac{1}{\Gamma}\frac{d\Gamma}{d\cos\theta
_{b,0}}=\frac{1}{2}[1+P\alpha\cos\theta _{b,0}] \label{eq:z-tb}
\end{equation}  
where $\alpha$ is the spin-analysing power of the $b$ quark
in the presence of anomalous $Wtb$ coupling $f_{2R}$. It includes
correction to the SM tree level value of $\alpha _b$. To second order in
$f_{2R}$, $\alpha$ can be written as 
\begin{align} 
\label{eq:alpb}
\alpha
&=-\left(\frac{\xi -2}{\xi+2}\right)+\operatorname{Re}(f_{2R})\left(\frac{8\sqrt{\xi}(\xi
-1)}{(\xi+2)^2}\right)\\ \nonumber &+|f_{2R}|^2\left(\frac{4(\xi-1)(\xi
^2-9\xi+2)}{(\xi+2)^3}\right)  
\end{align} 
Substituting
the values of $m_t=173.5 \mathrm{GeV}/\mathrm{c}^2$ and
$m_W=80.385\mathrm{GeV}/\mathrm{c}^2$ \cite{Beringer:1900zz}, we get the
numerical value of $\alpha$ as 
\begin{equation*}
\alpha=-0.399+1.425\operatorname{Re}(f_{2R})-0.903|f_{2R}|^2. 
\end{equation*} 
Since the
coefficiets of $f_{2R}$ and $f_{2R}^2$ are much greater than the
constant term in the above equation and the alternate terms differ in
their signs, cancellation between different terms can occur. Therefore
the effect of the anomalous coupling $f_{2R}$ on the $z$ distribution is
non-trivial. As an observable based on the ratio of the energy of the
top quark decay products (in the lab frame), $A_z$ can be used along
with $A_u$ to probe top-quark polarisation at large boosts. 

\begin{figure*}[t!]
\includegraphics[keepaspectratio=true,scale=0.35]{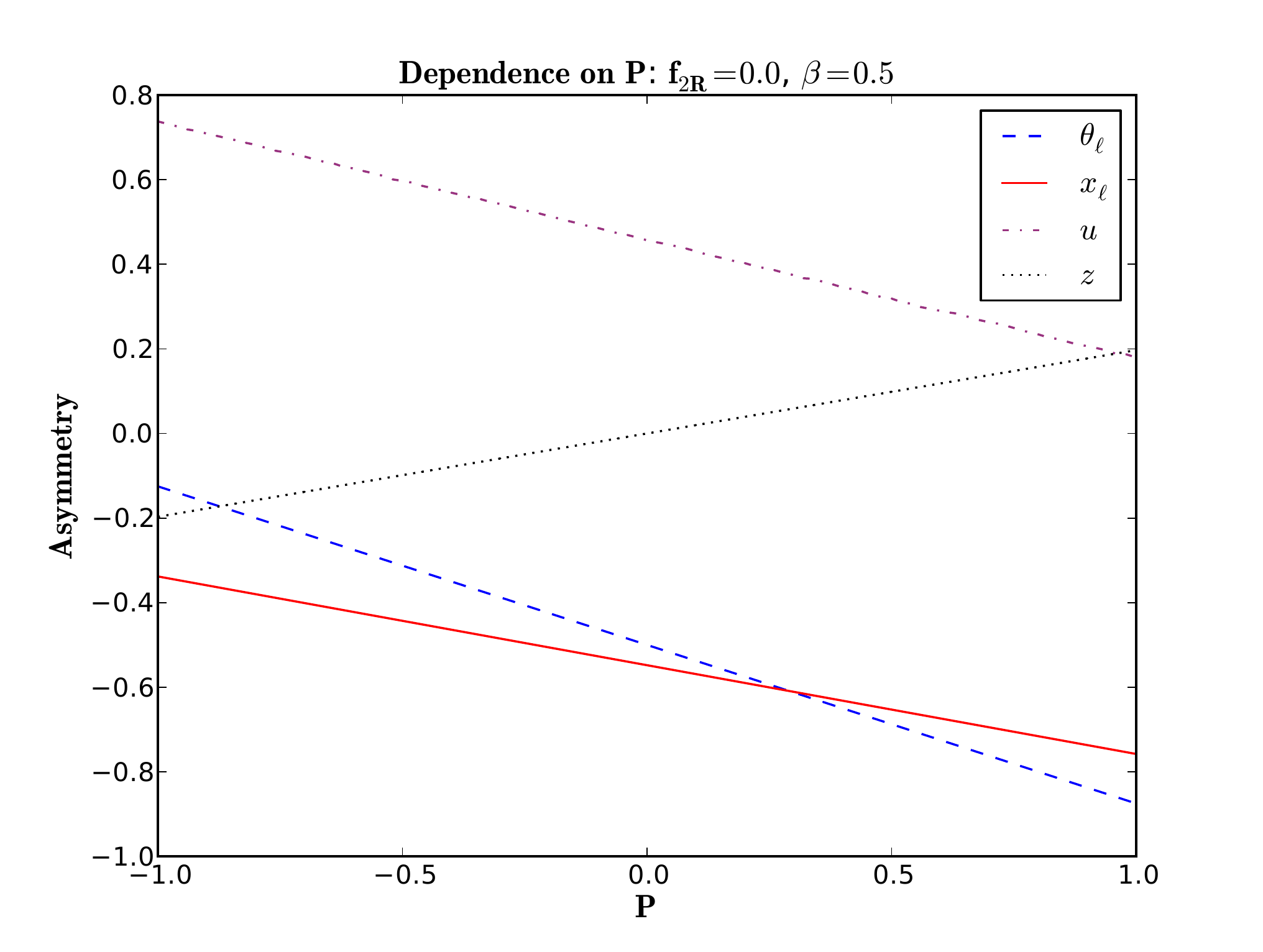}
\includegraphics[keepaspectratio=true,scale=0.35]{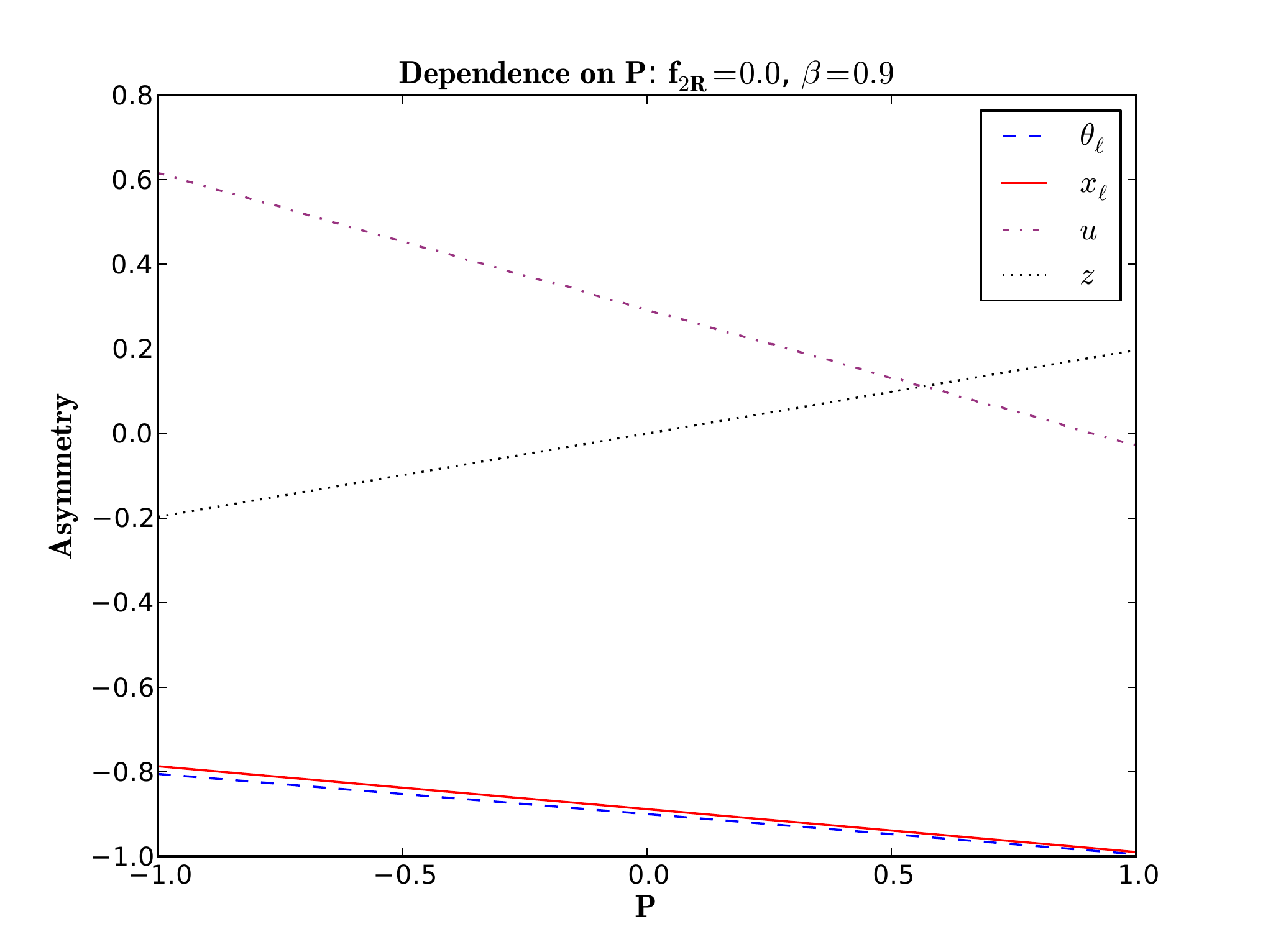}
\caption{Comparison of asymmetries in their dependence on the
polarisation of the top quark ($P$) in the lab frame. The boost factors
are $\beta =0.5$ (left) and $\beta=0.9$ (right) respectively.}
\label{fg:1} 
\end{figure*} 
\subsection{CP violation in the top decay}
 Here we note that using the asymmetries constructed for the $t$ and $\bar{t}$ decay, one can probe CP violation in the decay of top and anti-top assuming CP conservation in the production of top and anti-top quarks. As mentioned earlier, in the presence of CP conservation $f_{1L}=\bar{f}_{1L}$, $f_{1R}=\bar{f}_{1R}$, $f_{2L}=\bar{f}_{2R}$ and $f_{2R}=\bar{f}_{2L}$.  When the production process is CP-conserving, polarizations of the top ($P$) and the anti-top ($\bar{P}$) are related: $\bar{P}=-P$.  In this limit, the difference in the $u$-asymmetries of the top and the anti-top decay is proportional to $\operatorname{Re}(f_{2R})-\operatorname{Re}(\bar{f}_{2L})$ to linear order in the anomalous couplings (see eq. ~\ref{eq:11}). The coefficient of proportionality is a function of top polarization ($P$) and the boost $\beta$. Here $\bar{f}_{1L}$ is set to unity, $\bar{f}_{1R}$ and $\bar{f}_{2R}$ to zero. Similarly, the difference in $z$-asymmetries of the top and the anti-top is proportional to $\operatorname{Re}(f_{2R})-\operatorname{Re}(\bar{f}_{2L})$ with a factor $-4P\sqrt{\xi}(\xi -1)/(\xi +2)^2$ for  $\bar{f}_{1L}=1$ and $\bar{f}_{2R}=0=\bar{f}_{1R}$. Note that CP violation in decay necessarily requires an absorptive part in the amplitude and hence in any underlying theory it is expected that    $\operatorname{Re}(f_{2R})-\operatorname{Re}(\bar{f}_{2L})$ would be loop suppressed.

\subsection{Sensitivity of the asymmetries to
$P$ and $f_{2R}$}\label{ssec:kfrdep}  
\begin{figure*}[t!]

\includegraphics[keepaspectratio=true,scale=0.35]{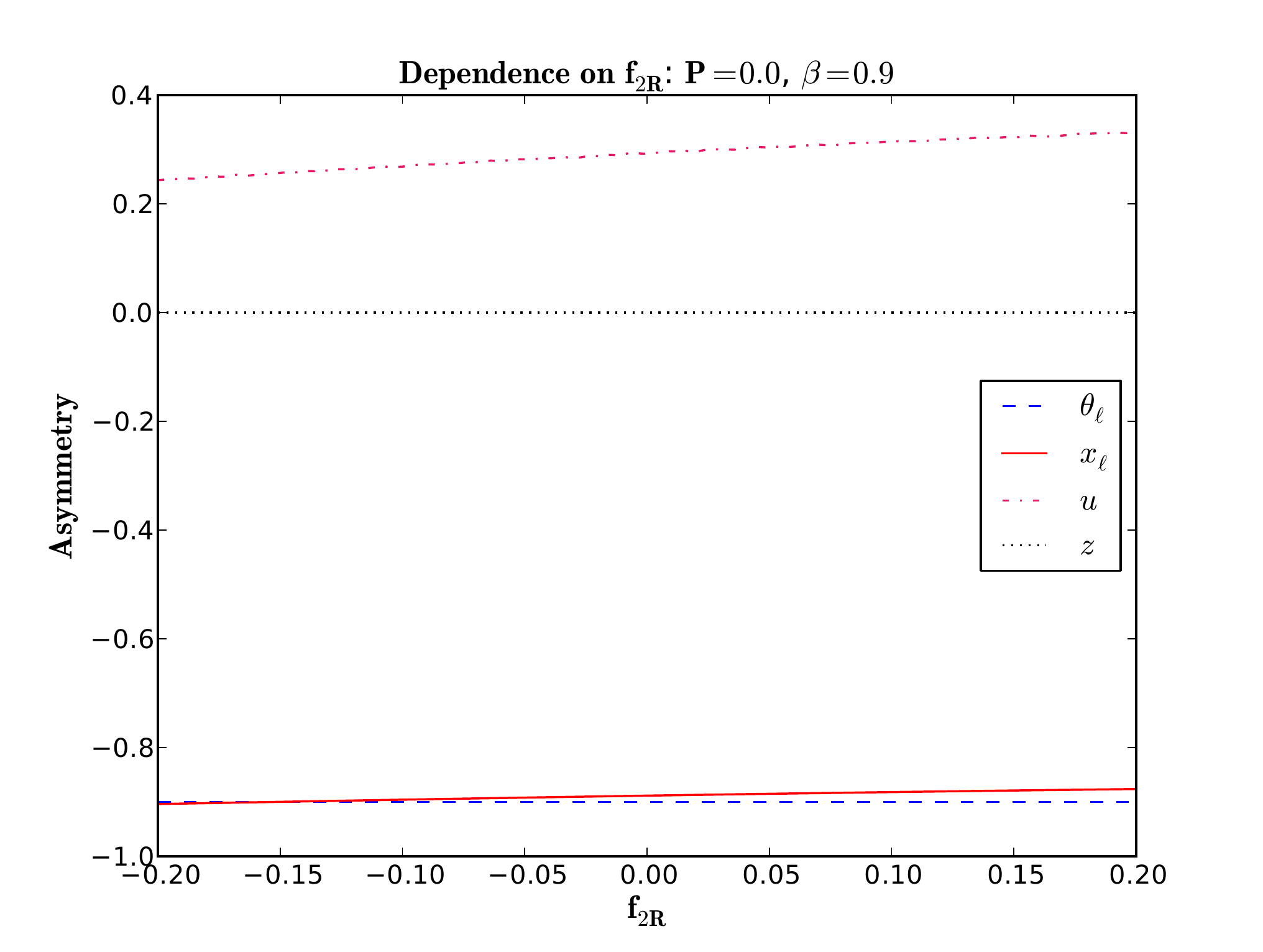}
\includegraphics[keepaspectratio=true,scale=0.35]{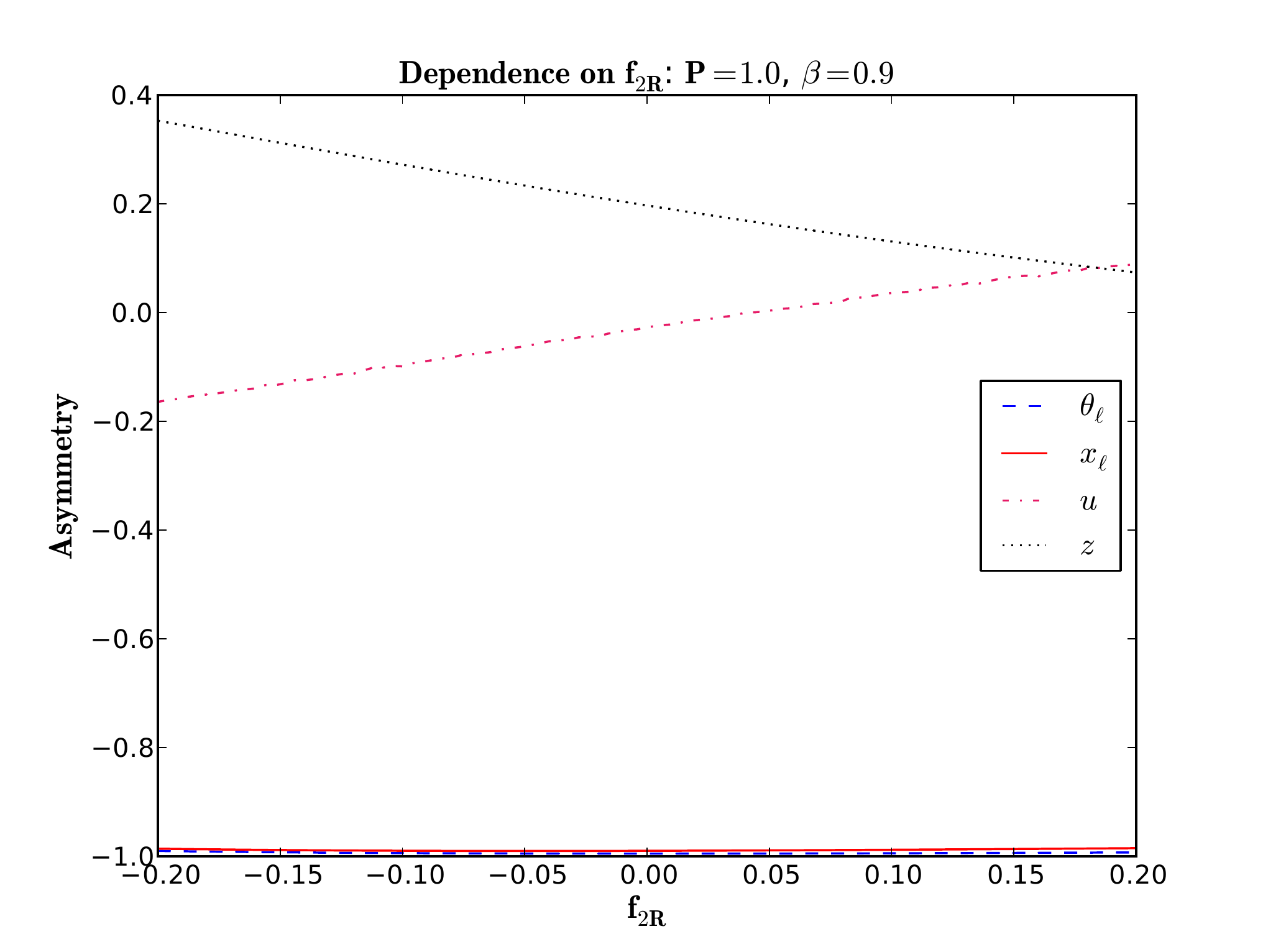}
\includegraphics[keepaspectratio=true,scale=0.35]{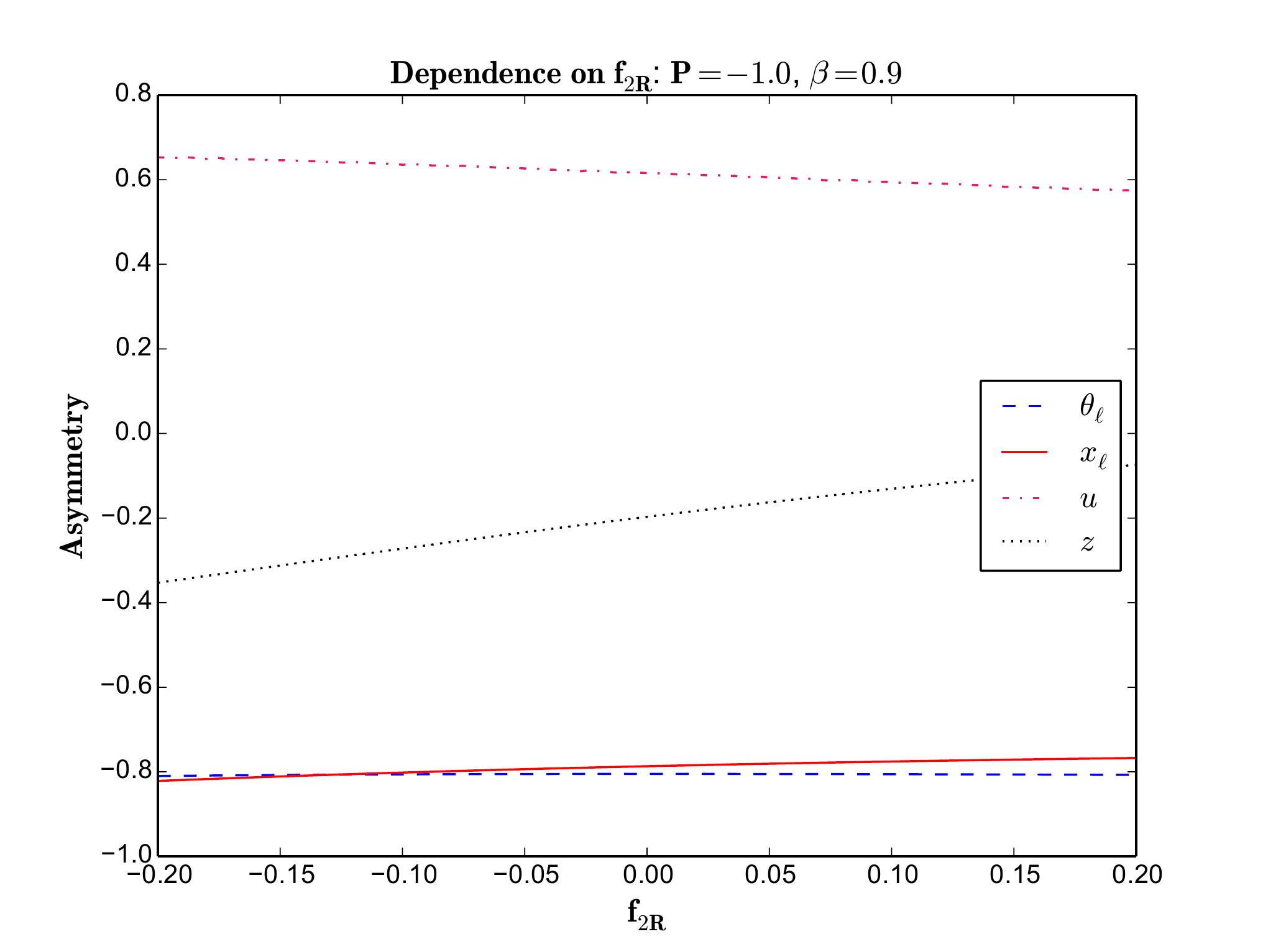}
\includegraphics[keepaspectratio=true,scale=0.35]{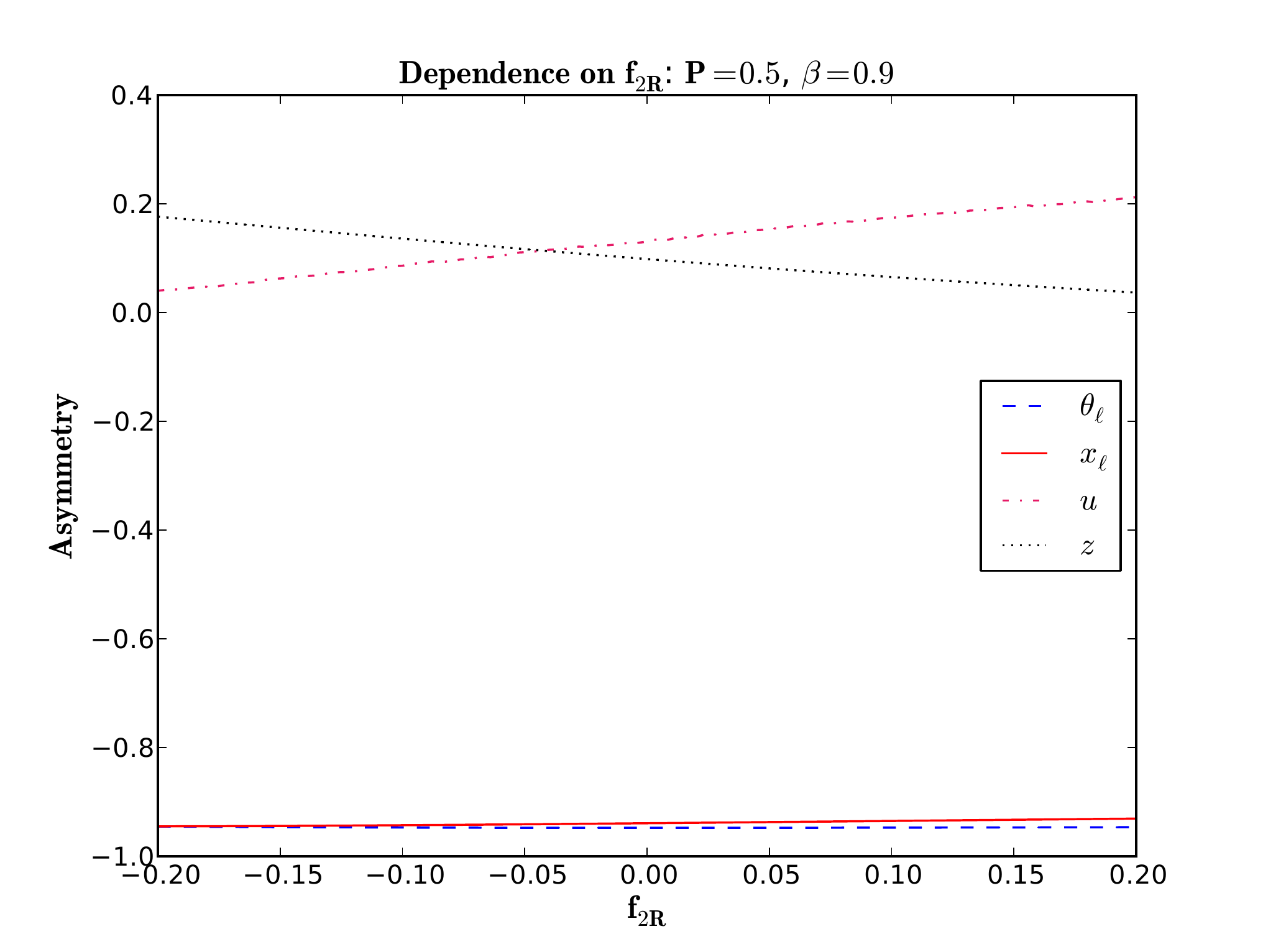}

\caption{Comparison of the $f_{2R}$ dependence of various
asymmetries for different values of $P$ and $f_{2R}$ for a boost factor
$\beta =0.9$. In each plot the $u$ asymmetry is given in dot-dashed
lines, the $x_{\ell}$ asymmetry in solid lines, the $\theta _{\ell}$
asymmetry in dashed lines and the $z$ asymmetry in dotted lines
respectively.}    
\label{fg:fr1} 
\end{figure*}

The
dependences of various asymmetries on the polarisation of the top quark
are compared in fig.~\ref{fg:1}. One can observe that for moderate values of boost $\beta
\sim 0.5$ all the four asymmetries are sensitive to the top polarisation
while for large values of boost, $A_u$ and $A_z$ are more
sensitive as compared to $A_{x_{\ell}}$ and $A_{\theta _{\ell}}$. For very small 
values of boosts ($\beta\approx 0$), the angular asymmetry $A_{\theta _{\ell}}$
 has the highest sensitivity to the top-quark polarisation ($P$) due the
 fact that the charged lepton has the maximal spin-analysing power. 
$A_z$ follows $A_{\theta _{\ell}}$ in the sensitivity to 
$P$ for $\beta\approx 0$ as it is directly proportional to the 
spin-analysing power of the $b$ quark ($\alpha$). This 
is true as long as the value of the anomalous coupling $f_{2R}$ does 
not reduce the value of $|\alpha |$. From equation eq.~\ref{eq:alpb} 
or from the plots of $A_z$ in fig. ~\ref{fg:fr1}, one can easily 
see that the value of $\alpha$ monotonically increases with $f_{2R}$ 
in the range $[-0.2,0.2]$. Therefore as a measure of top quark 
polarisation, $A_z$ is better for negative values of $f_{2R}$ than 
for positive values. In a detailed comparison, for 
$\beta\approx 0$, $A_{z}$ turns out to be the second best
in the sensitivity to $P$, surpassed only by 
$A_{\theta _{\ell}}$.

An additional result 
of the comparison is that the sensitivity of $A_u$ to $P$ is 
higher than that of $A_{x_{\ell}}$ in general (see fig.~\ref{fg:1}).   

Regarding the sensitivity of the  asymmetries to $f_{2R}$, an
interesting interplay of the top-quark polarisation and the anomalous
coupling $f_{2R}$ can be seen from fig.~\ref{fg:fr1}. For large
boost values ($\beta \sim 1$) $A_{\theta _{\ell}}$ and $A_{x_{\ell}}$
have similar sensitivities to $f_{2R}$ (for small values of $f_{2R}$) which is clearly shown in
fig.~\ref{fg:fr1}. In the case of $A_{z}$, a non-zero polarisation of
the top quark ($P$) is necessary to probe the anomalous coupling $f_{2R}$
since the asymmetry is directly proportional to $P$ (subsection
~\ref{ssec:1}). Moreover $A_z$ is independent of $\beta$ as mentioned 
above (subsection ~\ref{ssec:4}). This makes $A_z$ a suitable probe 
of $f_{2R}$ for all values of beta as long as $P\neq 0$. When $P=0$,
 $A_u$ and $A_{x_{\ell}}$ can be used to measure $f_{2R}$ although the 
sensitivity of $A_{x_{\ell}}$ to $f_{2R}$ (for small $f_{2R}$) is low
 at large values of boosts (fig. ~\ref{fg:fr1}). The angular asymmetry 
is not suitable to measure $f_{2R}$ (as long as  $f_{2R}$ is small) for 
any value of the boost as the spin analysing power $\alpha _{\ell}$ is 
insensitive to $f_{2R}$ (subsection ~\ref{ssec:1}). From  
fig. ~\ref{fg:fr1} one can say that for large boosts, $A_{u}$ 
can be used to measure $f_{2R}$ irrespective of the top-quark 
polarisation $P$. Therefore $A_u$ is the only observable that 
can be used to measure $f_{2R}$ at large boosts, for any production 
mechanism of the top quark. However, note that a comparison of
the asymmetries for their sensitivities to both the polarisation $P$
and the anomalous coupling $f_{2R}$, in a realistic experimental set up,
requires a careful study of various detector effects on the measurement 
of asymmetries, and is beyond the
scope of the present work (see, for example \cite{Papaefstathiou:2011kd}).

\section{Constraining $P$ and $f_{2R}$ simultaneously}\label{sec:2} 
When $f_{2R}\neq 0$ in the
$Wtb$ vertex, the asymmetries considered above are
affected by the anomalous coupling $f_{2R}$ along with the polarisation
$P$
of the top. Therefore with the measurement of an asymmetry one
constrains a region in the two-dimensional $P$-$f_{2R}$ plane. In this
section we compare the asymmetries based on the region each one
constrains in the $P$-$f_{2R}$ plane assuming a plausible set of values of
asymmetries expected to be measured at the LHC. We also discuss
combining these asymmetries in a $\chi ^2$-based analysis.
\subsection{Method of analysis} 
We assume that the statistical error associated with the measurement of an asymmetry $A$ is given by
\begin{equation}\label{eq:15} 
\Delta A=\frac{1}{\sqrt{N}}\sqrt{1-A^2}
\end{equation} 
where $N$ is the number of top quarks in the sample of
measurement. Given the fact that experimentally measured observables have also systematic uncertainties coming from various sources such as missing higher order corrections to the theoretical predictions, uncertainties in the parton distribution functions, we need to include in $\Delta A$  an estimate of the systematic uncertainties associated with the asymmetry $A$.  The total uncertainty  on $A$ after including both the statistical and systematic uncertainties is given by 
\begin{equation}\label{eq:n15}
\Delta A=\sqrt{\frac{(1-A^2)}{N}+\frac{\epsilon ^2}{2}(1-A^2)^2}
\end{equation}
where $\epsilon$ is the fractional systematic uncertainty in the top production cross section at the LHC at $\sqrt{s}=7$ TeV.  The number $N$  appearing in the equations above is estimated from the expected number of top
quarks produced at the LHC from heavy resonances with invariant masses
of $O(\mathrm{TeV})$ decaying into a top pair.  
Based on the estimated differential cross section for top-pair
production
calculated in QCD \cite{Ahrens:2010zv} for the LHC at
$\sqrt{s}=7$ $\mathrm{TeV}$ in the invariant-mass window of $1.0$
$\mathrm{TeV}$ and $1.2$ $\mathrm{TeV}$,  we take the number of top quarks
as $N=1.47\times 10^4$ for an integrated luminosity of $100$
$\mathrm{fb}^{-1}$. This number is obtained under the assumption that the
top quark decays semileptonically $t\rightarrow b\bar{\ell}\nu$ with
$\ell=e,\mu$ and the anti-top decays hadronically.   Theoretically  an
asymmetry is a function of $P$ and $f_{2R}$ and the factor $\beta$ is
close to unity as we consider only those top quarks which are highly
boosted  in the lab frame. In fact, the boost values of the top 
quarks produced in the above-mentioned 
invariant-mass window are in the range 0.94 to 0.96. As
we intend to keep our analysis a qualitative one, we fix $\beta$ to $0.9$.
\begin{figure*}[t!]
\includegraphics[keepaspectratio=true,scale=0.35]{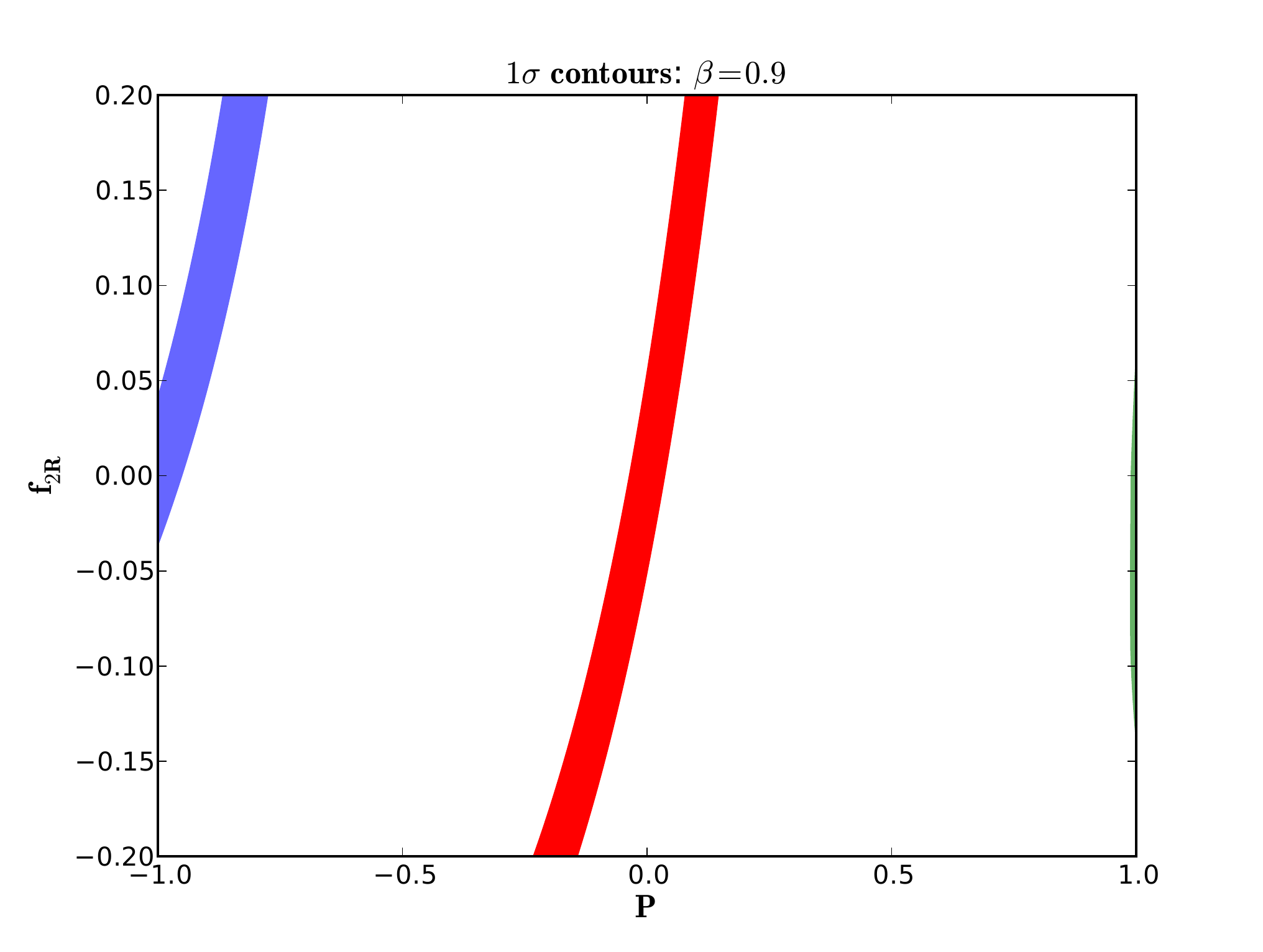}
\includegraphics[keepaspectratio=true,scale=0.35]{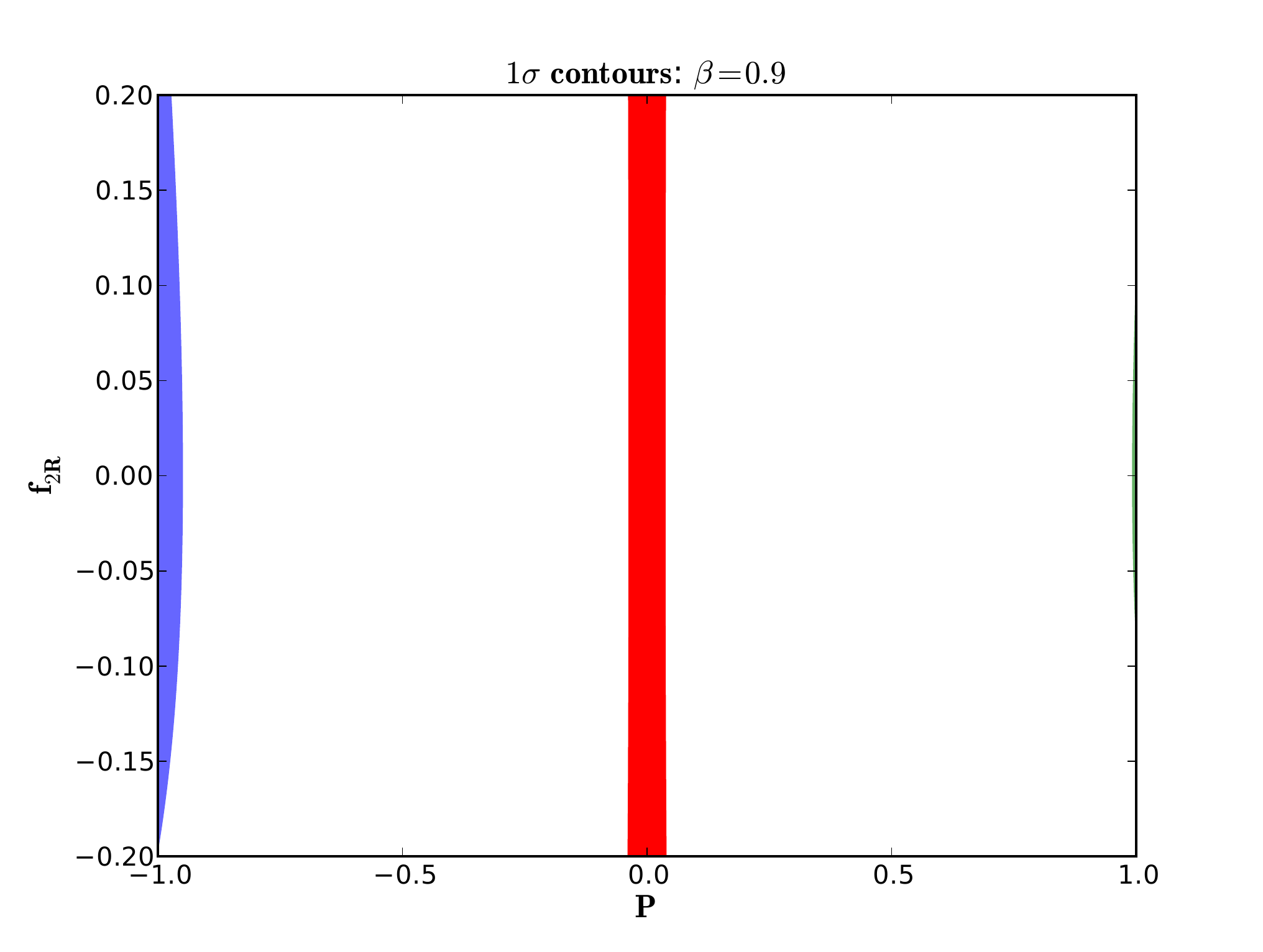}
\includegraphics[keepaspectratio=true,scale=0.35]{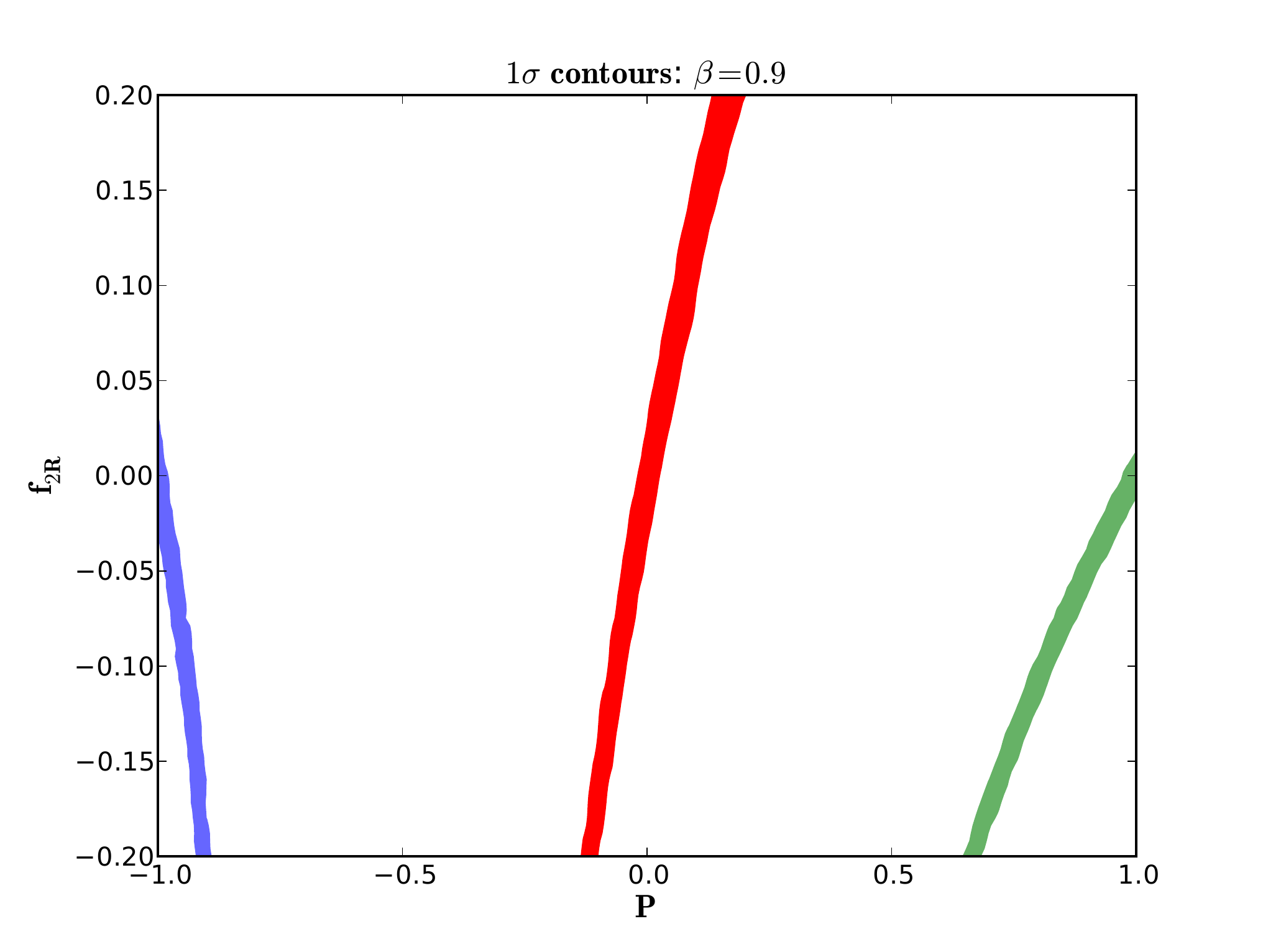}
\includegraphics[keepaspectratio=true,scale=0.35]{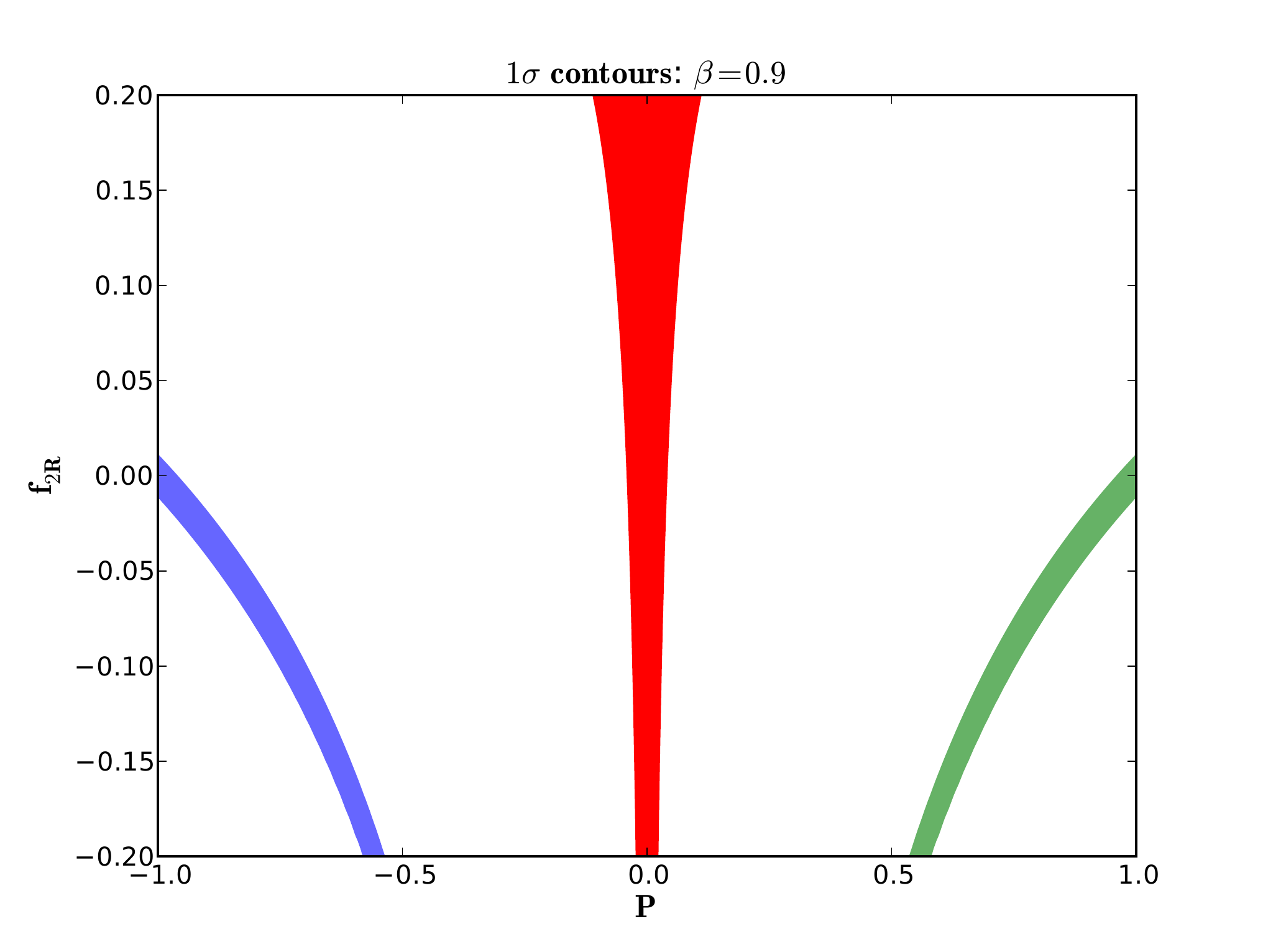}
 \caption{Comparison of regions with significance less than
or equal to 1 for different asymmetries defined in the lab frame: $A_{x
_{\ell}}$(upper left), $A_{\theta _{\ell}}$(upper right), $A_{u}$(bottom
left) and $A_z$(bottom right). The regions shaded in blue(very light),
red(dark), green(light) correspond to the ``true'' values
$P_0=-1.0,f_{2R0}=0.0$, $P_0=0.0,f_{2R0}=0.0$ and $P_0=1.0$,
$f_{2R0}=0.0$ respectively. The boost factor $\beta$ is set to $0.9$. In these figures 
only statistical uncertainties are assumed for the asymmetries.}\label{fg:2}
\end{figure*} 

Suppose that an experimental measurement of $A$ corresponds to a true
value $(P_0,f_{2R0})$ of the parameters $P$ and $f_{2R}$. This
measurement corresponds to an unknown point $(P_0,f_{2R0})$ in the
$P$-$f_{2R}$ plane. We define a region of siginificance $f$ around the
point $(P_0,f_{2R0})$ as the region where the value of the asymmetry
$A(P,f_{2R})$ is indistinguishable from the experimental value
$A_{\mathrm{exp}}$ to within $f$ times the error in the measurement
$\Delta A$. In other words, 
\begin{equation}\label{eq:16}
\frac{|A_{\mathrm{exp}}-A(P,f_{2R})|}{\Delta A_{\mathrm{exp}}}\leq f.
\end{equation} 
Since our purpose in this paper is to demonstrate the use
of asymmetries, we choose a value for $P_0$ and $f_{2R0}$; evaluate
$\Delta A_{\mathrm{exp}}$ using eq.~\ref{eq:15} keeping only the statistical uncertainties
and $A$ from the expressions of the corresponding
distributions derived in the previous section. 
The results are shown in fig. ~\ref{fg:2}.

\subsection{$\chi ^2$ analysis} We combine three of the four asymmetries
 to make a $\chi ^2$ statistic. One could combine all the four asymmetries to form a $\chi ^2$ statistic. We have not considered such a combination in our analysis. This is because we consider, in our analysis, the case of highly boosted top quarks where effectively only two asymmetries $A_u$ and  $A_z$ are sensitive to both $P$ and $f_{2R}$.  Moreover, in the case of  $P_0=0.0$ and $f_{2R0}=0.0$, one can easily see from fig.~\ref{fg:2} that the asymmetries  $A_{\theta _{\ell}}$ and $A_{z}$ are similar in their abilty to constrain $f_{2R}$. Similarly, in the case of $P_0=-1.0$ and $f_{2R0}=0.0$ , the bounds on $f_{2R}$ are primarily due to $A_z$ and $A_u$ which can be seen from fig.~\ref{fg:2}. The asymmetries $A_{\theta _{\ell}}$ and $A_{x_{\ell}}$ are relatively poor in constraining $f_{2R}$. In both cases either one of  $A_{x_{\ell}}$ and $A_{\theta _{\ell}}$ is sufficient to constrain $P$  (see fig. ~\ref{fg:2}). Hence inclusion all the four asymmetries at a time in the $\chi ^2$ analysis does not improve the best bounds obtained in the current analysis.

 There are four ways in which three of the
asymmetries $A_{x_{\ell}},A_{\theta _{\ell}},A_u,A_z$ can be combined.
We discuss each of the combination. We assume that the asymmetries are measured independently and their errors are given according to either eq.~\ref{eq:n15} or eq.~\ref{eq:15} depending on whether the systematic uncertainties are included or not. 
The $\chi ^2$ is defined by
\begin{equation}\label{eq:17} 
\chi ^2 =\sum
_{i}\left(\frac{A_{\mathrm{exp},i}-A_i(P,f_{2R})}{\Delta
A_{\mathrm{exp},i}}\right)^2 
\end{equation} 
where $i=x_{\ell},\theta_{\ell},u,z$.  

 Since our purpose is to demonstrate the utility of
combining asymmetries, we calculate $A_{\mathrm{exp}}$ for a ``true''
value of $P$ and $f_{2R}$ i.e $P_0$ and $f_{2R0}$ and evaluate  $\Delta A_{\mathrm{exp}}$ for two cases. In the first case only statistical uncertainties are included in $\Delta A_{\mathrm{exp}}$ using eq.~\ref{eq:15}. In the second case the systematic uncertainties are also included in $\Delta A_{\mathrm{exp}}$ as given in eq.~\ref{eq:n15}. 

We give contours of $\Delta\chi^2$ values $2.30$ and $5.99$ corresponding to $68.3\%$ and $95\%$ confidence level (C.L) (for 2 degrees of freedom) respectively for both cases. As in the
previous section we set $\beta=0.9$ and use the same number of events
$N$.  
\begin{figure*}[t!]

\includegraphics[keepaspectratio=true,scale=0.35]{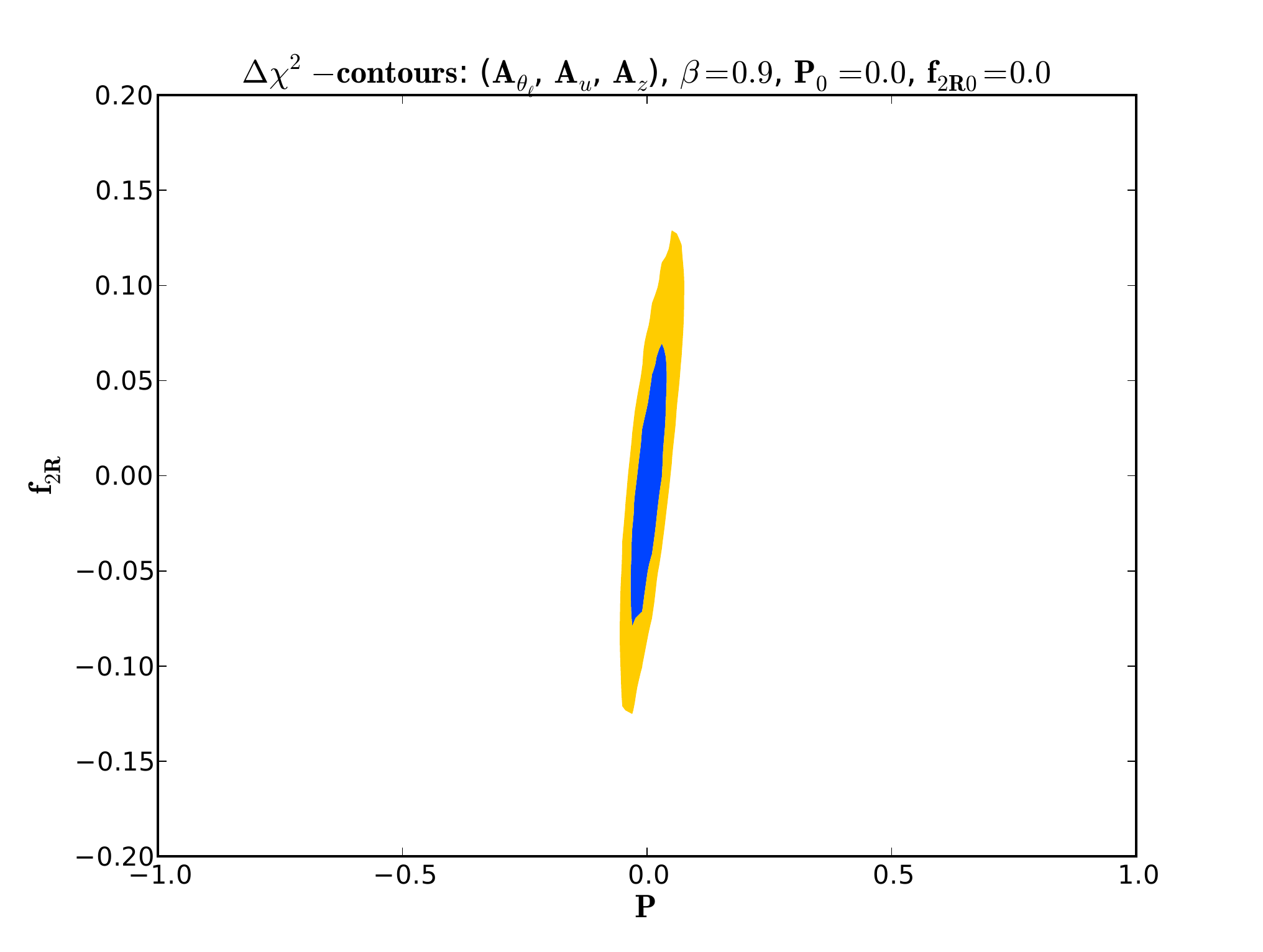}
\includegraphics[keepaspectratio=true,scale=0.35]{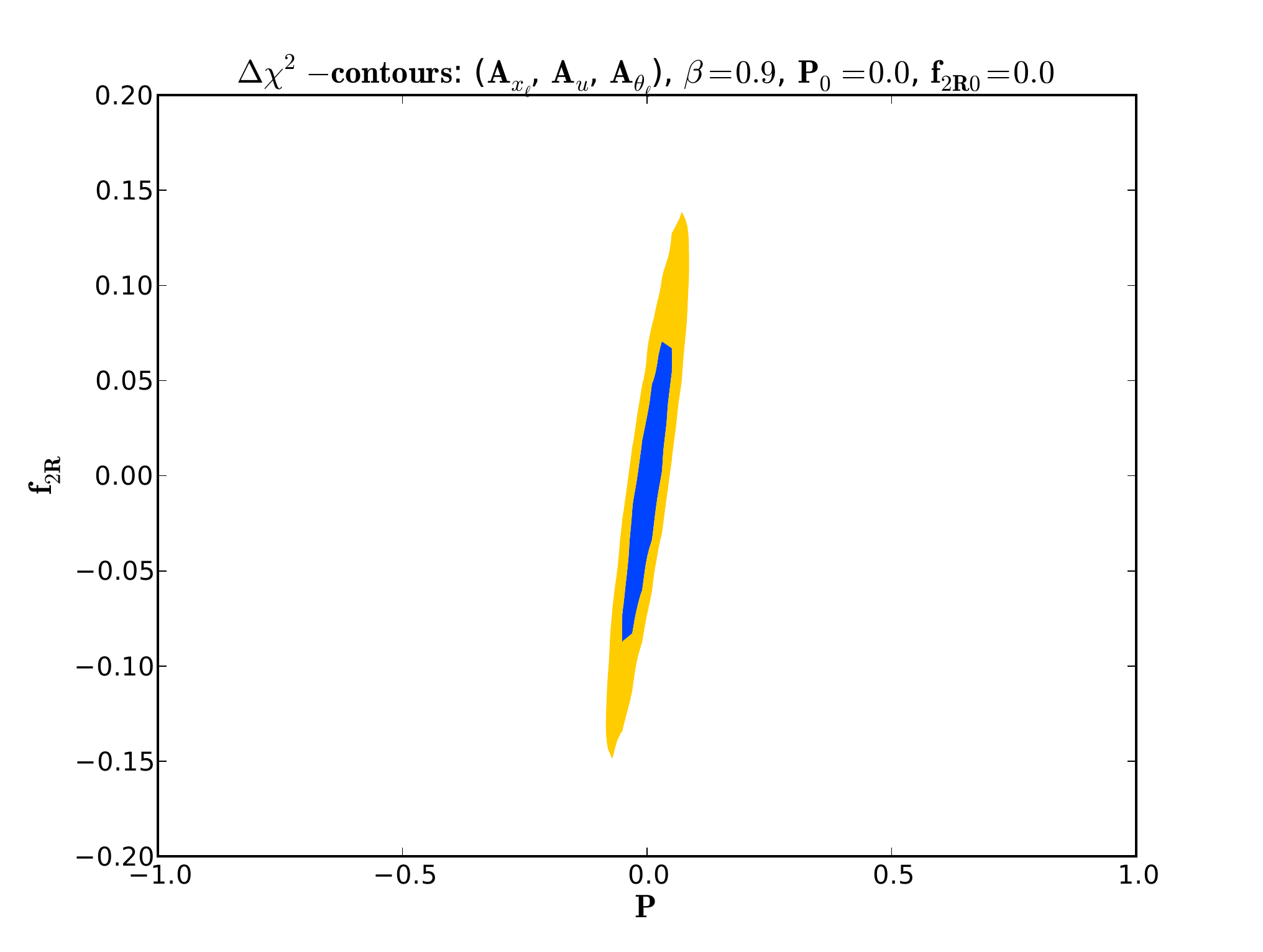}
\includegraphics[keepaspectratio=true,scale=0.35]{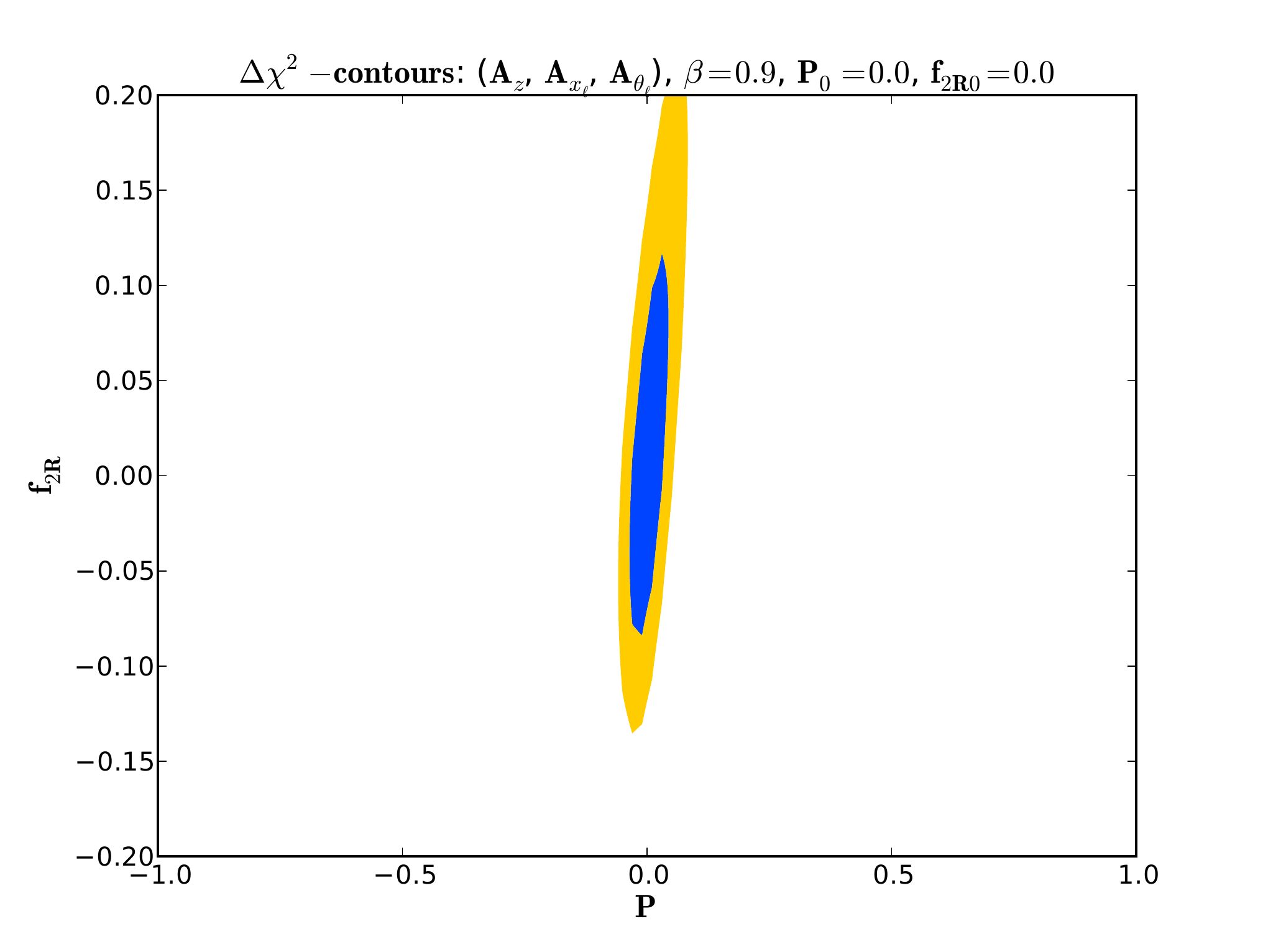}
\includegraphics[keepaspectratio=true,scale=0.35]{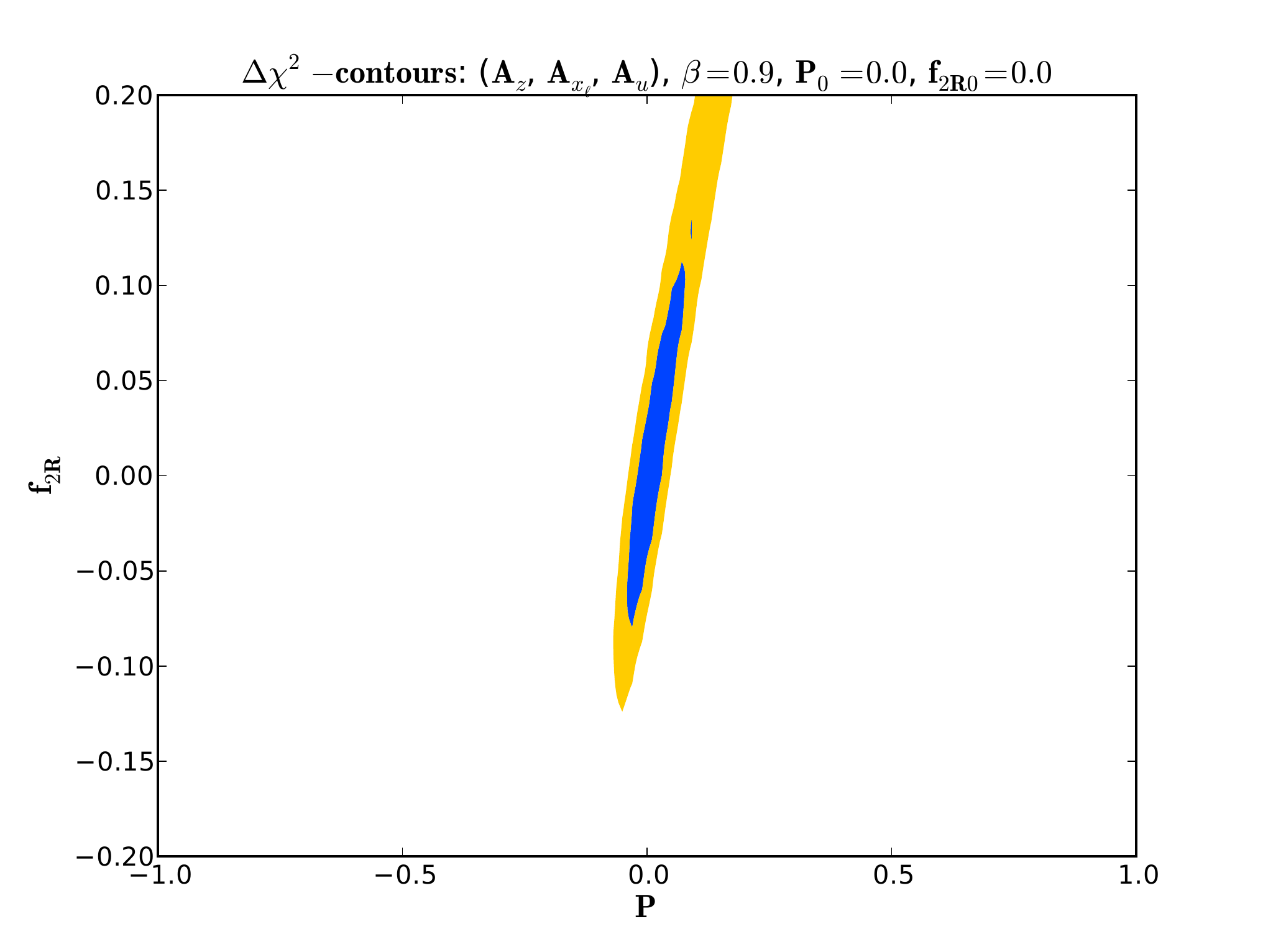}
\caption{Contours of $\Delta \chi ^2$ corresponding to
$68.27\%$ (blue/darker) and $95\%$ C.L (yellow/lighter) respectively for
four combinations of asymmetries:$A_{\theta _{\ell}},A_u,A_z$(top left),
$A_{x_{\ell}},A_u,A_{\theta _{\ell}}$(top right),
$A_z,A_{x_{\ell}},A_{\theta _{\ell}}$(bottom left) and
$A_z,A_{x_{\ell}},A_u$(bottom right). The boost factor is set to
$\beta=0.9$. The ``true'' values of $P$ and $f_{2R}$ are $P_0=0.0$ and
$f_{2R0}=0.0$ respectively. These contours are for the case where only statistical uncertainties are assumed for the asymmetries.} 
\label{fg:3} 
\end{figure*}

\begin{figure*}[t!]

\includegraphics[keepaspectratio=true,scale=0.35]{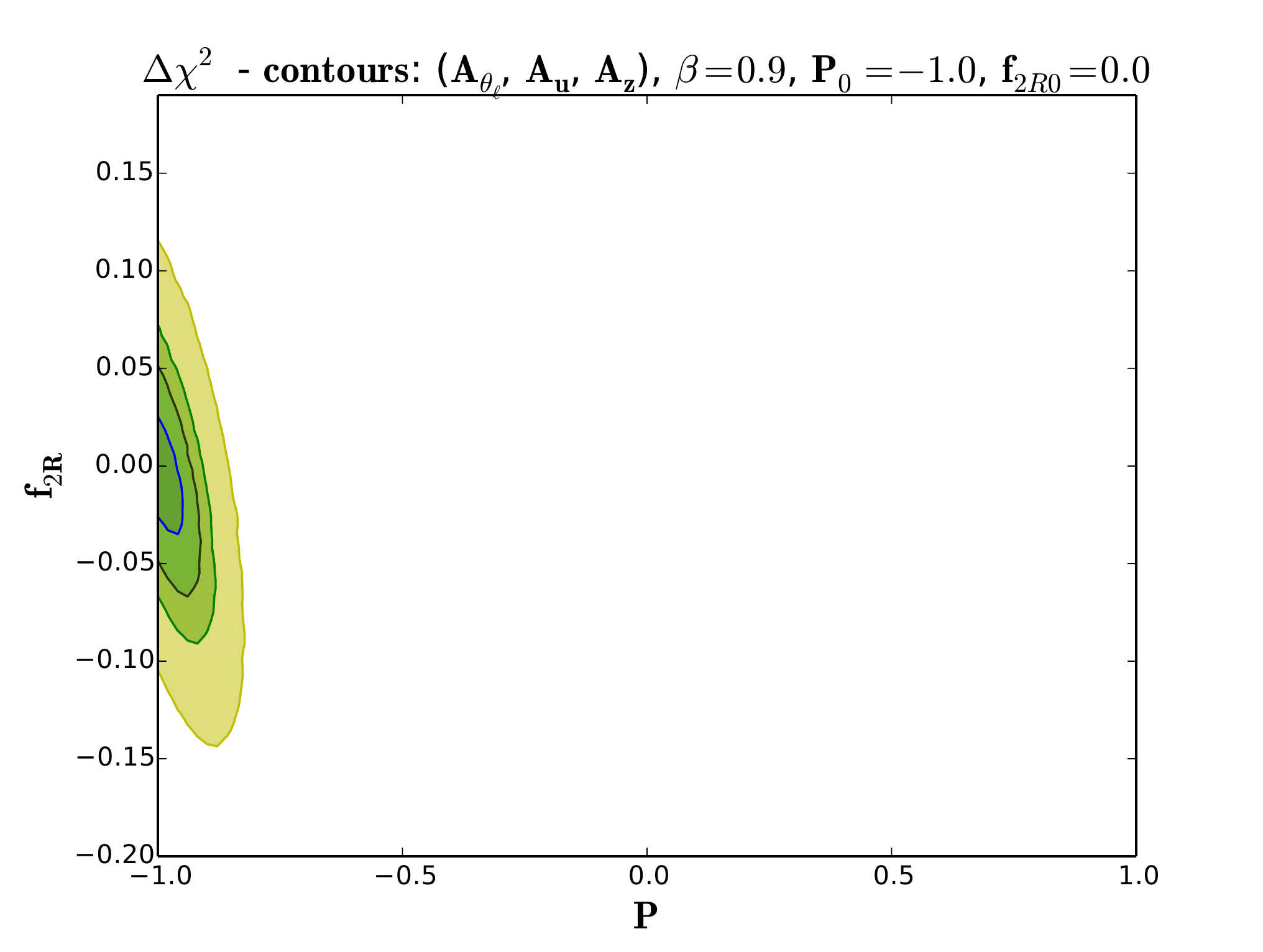}
\includegraphics[keepaspectratio=true,scale=0.35]{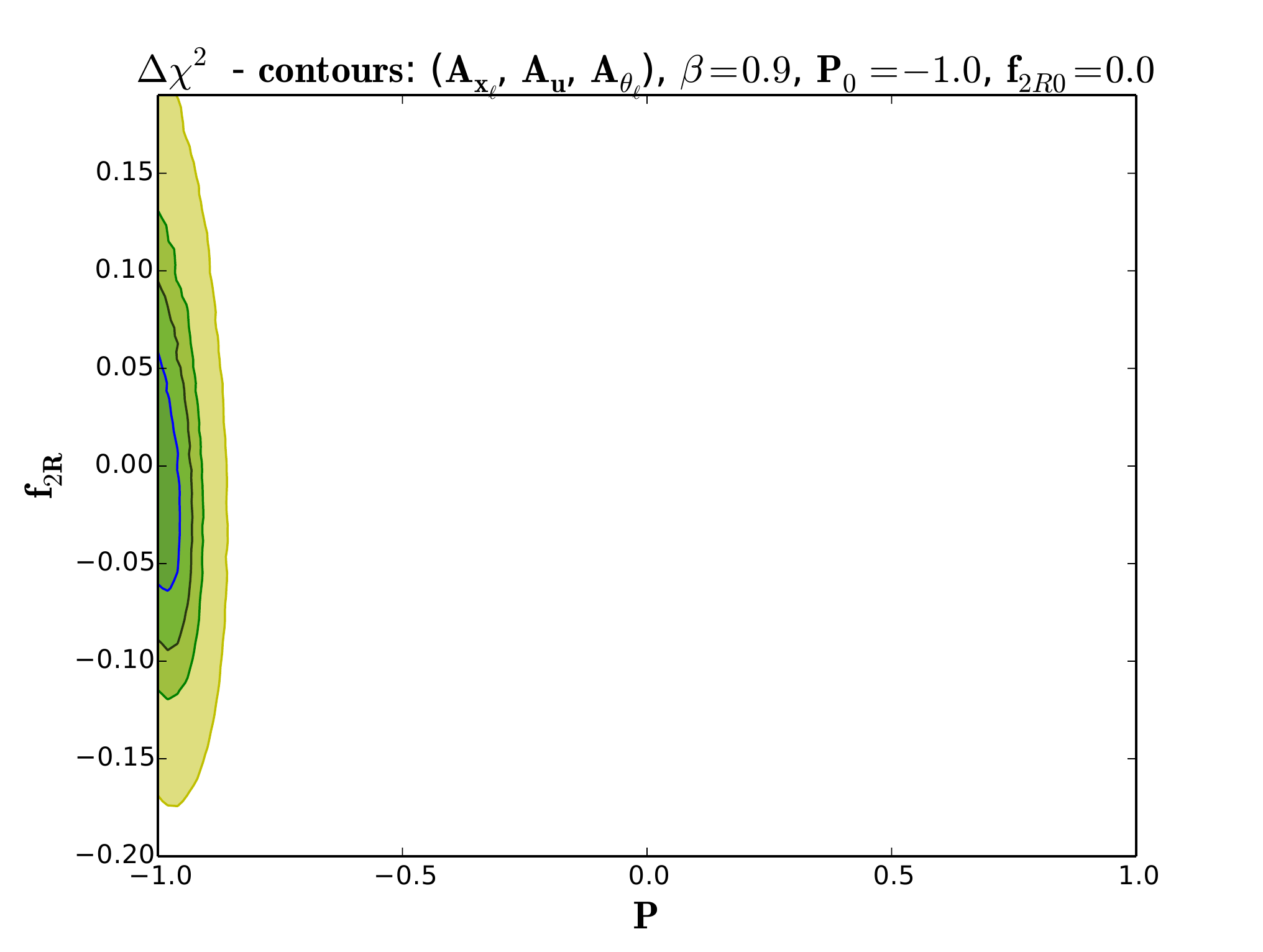}
\includegraphics[keepaspectratio=true,scale=0.35]{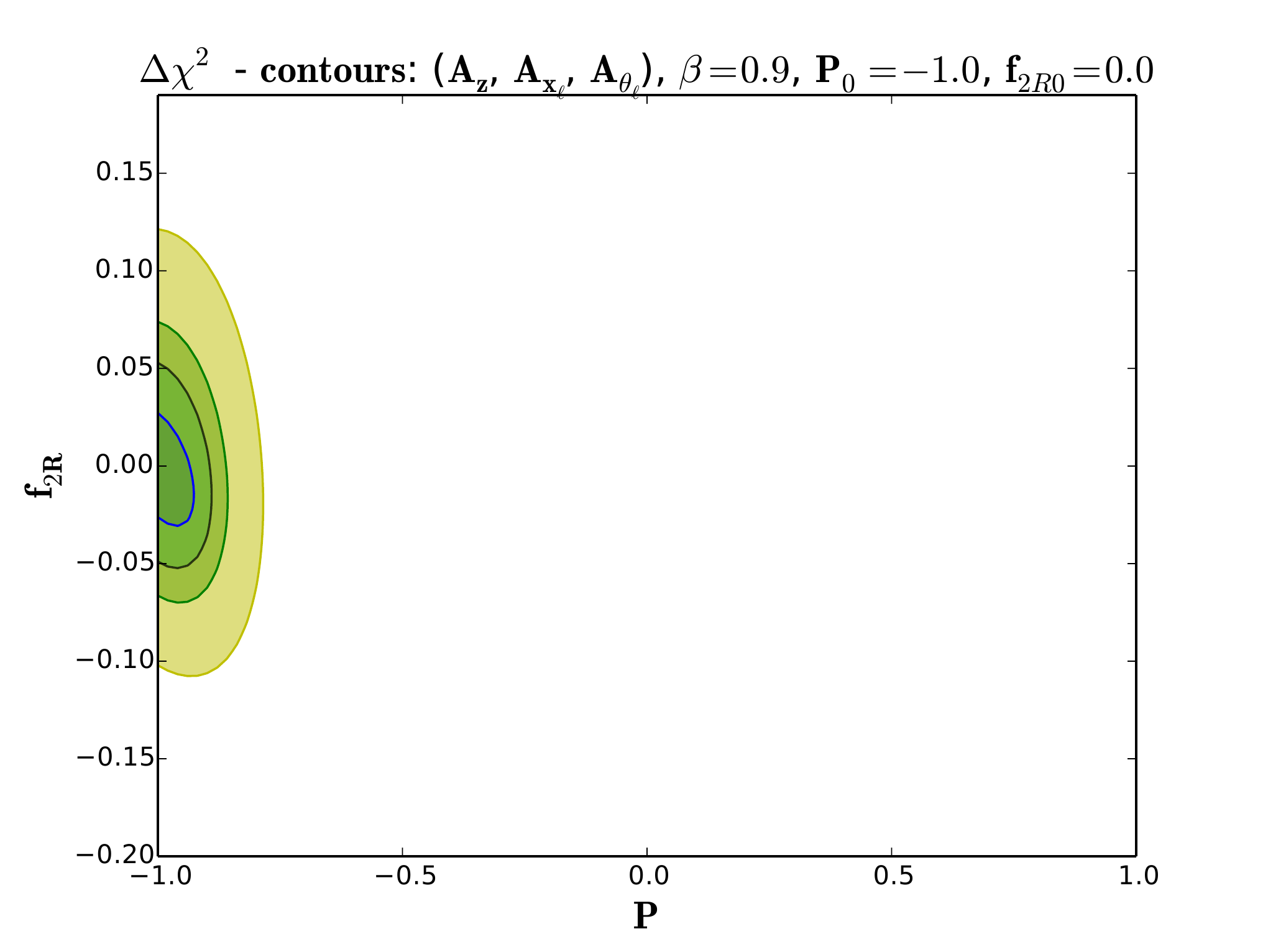}
\includegraphics[keepaspectratio=true,scale=0.35]{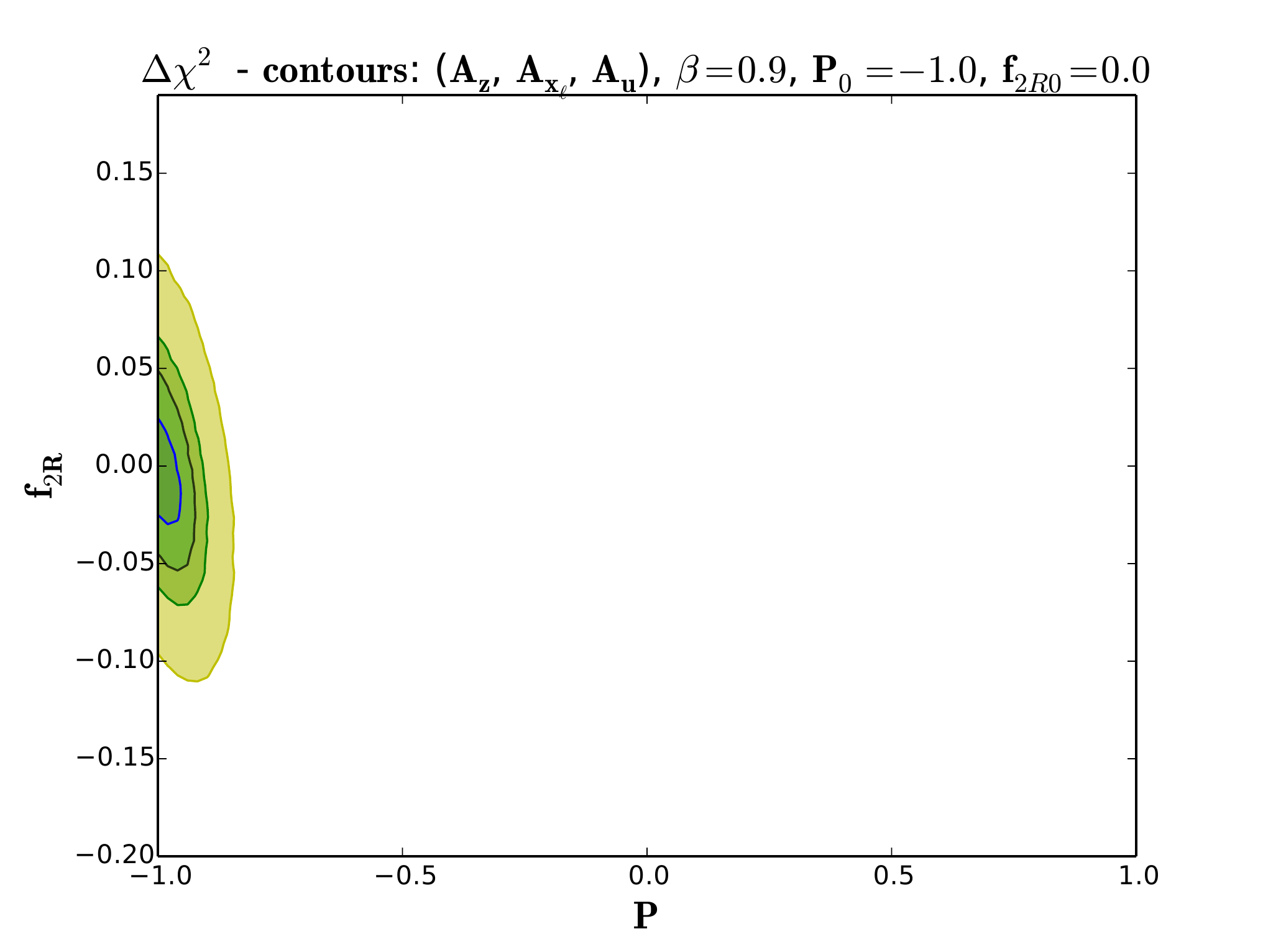}
\caption{ Contours of $\Delta \chi ^2$ corresponding to  $95\%$ C.L are given for four values of the systematic uncertainty parameter $\epsilon$ for each of the 
four combinations of asymmetries:$A_{\theta _{\ell}},A_u,A_z$(top left),
$A_{x_{\ell}},A_u,A_{\theta _{\ell}}$(top right),
$A_z,A_{x_{\ell}},A_{\theta _{\ell}}$(bottom left) and
$A_z,A_{x_{\ell}},A_u$(bottom right). The systematic uncertainties associated with the asymmetries are calculated according to eq.~\ref{eq:n15}. In each figure, the darker to lighter contours correspond to $\epsilon=0.0$, $0.02$, $0.03$, $0.05$ respectively. The boost factor is set to
$\beta=0.9$. The ``true'' values of $P$ and $f_{2R}$ are $P_0=-1.0$ and
$f_{2R0}=0.0$ respectively.} 
\label{fg:3c} 
\end{figure*}

\begin{table} 
\begin{tabular}{|c|c|c|} 
\hline combination & $1\sigma$ &
$2\sigma$\\ \hline $A_z,A_{x_{\ell}},A_{\theta _{\ell}}$ & $-0.96$ &
$-0.94$\\ $A_z,A_u,A_{x_{\ell}}$ & $-0.98$ & $-0.96$\\
$A_{x_{\ell}},A_u,A_{\theta _{\ell}}$ & $-0.98$ & $-0.96$\\ $A_{\theta
_{\ell}},A_u,A_z$ & $-0.97$ & $-0.95$\\ \hline 
\end{tabular} 
\caption{The
upper limts on the polarisation of the top ($P$) corresponding to
$f_{2R}=0$ from a $\Delta\chi ^2$-analysis. The true values of the
parameters are $P_0=-1.0$ and $f_{2R0}=0.0$. The lower limit on $P$ is
the physical boundary $P=-1.0$. Only statistical uncertainties are assumed for the asymmetries.} 
\label{tab:1} 
\end{table} 
Fig. ~\ref{fg:3} shows the $\Delta \chi ^2 $ contours for four different
combinations of asymmetries for $\beta=0.9$ keeping only the statistical uncertainties. The effects of including systematic uncertainties of the asymmetries in the $\chi ^2$ analysis are given later in the text. Table ~\ref{tab:1}
 gives the upper bound obtained on $P$ when
$f_{2R}=0.0$, for different combinations of asymmetries, for $\beta=0.9$.
When the true value of $P$ and $f_{2R}$ are $P_0=0,f_{2R0}=0$ the
combination of $A_{\theta _{\ell}}$,$A_u$ and $A_z$ and
$A_{x_{\ell}},A_u,A_{\theta _{\ell}}$ are better in constraining both
$P$ and $f_{2R}$ than the other two combinations. 

  \subsection{Limits on
$f_{2R}$} 

\begin{table}
\begin{tabular}{|c|c|c|} 
\hline combination & $1\sigma$ &
$2\sigma$\\
\hline ($A_{\theta _{\ell}}$, $A_{u}$, $A_{z}$) &
$[-0.079,0.069]$ & $[-0.125,0.129]$\\ 
($A_{x_{\ell}}$, $A_u$, $A_{\theta
_{\ell}}$) & $[-0.087,0.07]$ & $[-0.149,0.138]$\\ 
($A_{z}$, $A_{x_{\ell}}$,
$A_{\theta _{\ell}}$) & $[-0.083,0.116]$ & $[-0.135,0.2]$\\ 
($A_{z}$,
$A_{x_{\ell}}$, $A_{u}$) & $[-0.079,0.112]$ & $[-0.124,0.2]$\\
\hline
\end{tabular}

\caption{Limits on $f_{2R}$ at $1\sigma$ and $2\sigma$ level
corresponding to the polarisation $P=0$. Only statistical uncertainties are assumed for the asymmetries.} \label{tab:2} 
\end{table}

\begin{table} 
\begin{tabular}{|c|c|c|} 
\hline combination & $1\sigma$ &
$2\sigma$\\
\hline ($A_{\theta _{\ell}}$, $A_{u}$, $A_{z}$) &
$[-0.017,0.013]$ & $[-0.027,0.024]$\\ 
($A_{x_{\ell}}$, $A_u$, $A_{\theta
_{\ell}}$) & $[-0.039,0.012]$ & $[-0.064,0.059]$\\ 
($A_{z}$, $A_{x_{\ell}}$,
$A_{\theta _{\ell}}$) & $[-0.019,0.016]$ & $[-0.031,0.027]$\\ 
($A_{z}$,
$A_{x_{\ell}}$, $A_{u}$) & $[-0.017,0.006]$ & $[-0.029,0.018]$\\
\hline
\end{tabular}

\caption{Limits on $f_{2R}$ at $1\sigma$ and $2\sigma$ level
corresponding to the polarisation $P=-1.0$. Only statistical uncertainties are assumed for the asymmetries.} \label{tab:3} 
\end{table}

In the table ~\ref{tab:2} and ~\ref{tab:3} we summarise the
limits obtained on the anomalous coupling $f_{2R}$ for two values of 
polarisation $P=0$ and $P=-1.0$ keeping only statistical uncertainties. The best $1\sigma$ limits on $f_{2R}$ obtained
 in our analysis assuming
the top polarisation to be zero is $[-0.079,0.069]$. The sensitivity increases considerably
if, for example, the expected polarization of the top would be $-1.0$. The corresponding limit on $f_{2R}$ is:
$[-0.017,0.006]$. Now we discuss the results after including systematic uncertainties of asymmetries using eq.~\ref{eq:n15} in eq.~\ref{eq:17}. The main effect of such inclusion is that the limits on $f_{2R}$ and $P$ become weaker after the inclusion of systematic uncertainties.
In particular, for $\epsilon\gtrsim 1\%$ the $\chi ^2$ statistic does not constrain $f_{2R}$ to between $[-0.2,0.2]$ in the case of $P_0=0.0$. In this case, one may need to use methods such as multivariate analysis, fit to the shape of the distributions, etc. to constrain $P$ and $f_{2R}$ simultaneously. However, for large values of $P_0$ our observables are still sentitive to both $P$ and $f_{2R}$ for values of $\epsilon$ upto $5\%$ as can be seen from fig.~\ref{fg:3c}.  A detailed analysis to estimate the systematic uncertainty on the asymmetries would take into account the effects of hadronisation, finite detector resolution etc on the measurement of asymmetries. Such an analysis would be very useful in improving the bounds on $P_0$ and $f_{2R}$ compared to our simplified approach and is in progress.

Note that we have analyzed the case of $t \bar t$ pair production using  events with $t \bar t$ invariant masses
in the range $1.0$ $\mathrm{TeV}$ to $1.2$ $\mathrm{TeV}$ so as to analyse $t\bar{t}$ pairs possibly coming from a resonance. Even with an integrated luminosity
of 100 fb$^{-1}$ at $\sqrt{s}=7$ TeV LHC, the number of top events  in this analysis is
considerably lower compared to the number used in the analyses such as
\cite{AguilarSaavedra:2006fy} which uses $t\bar{t}$ events over the entire range of the invariant mass. Due to the lower statistics our limits on
$f_{2R}$ for the case of zero polarisation are weaker compared to the
limits obtained in \cite{AguilarSaavedra:2006fy}. But they are compatible
with the CMS measurement of $f_{2R}$:$0.07\pm 0.053
(\mathrm{stat})^{+0.081}_{-0.073}(\mathrm{syst})$ using 2.2 $fb^{-1}$ of
integrated luminosity at $\sqrt{s}=7$ TeV. Note that our results are compatible with the result obtained by  the CMS experiment with 2.2 fb$^{-1}$ luminosity (corresponding to  a number of events $N\sim O(10 ^4)$) and hence smaller number of $t \bar t$ events than the ATLAS analysis, using the $t \bar t$ events with invariant-masses over the whole allowed range. This gives us confidence that the various limits indicated in this report are representative of what can be achieved in a real analysis. However,  with this luminosity the observables are not sensitive to the contribution of the SM higher order corrections to the anomalous couplings, including $f_{2R}$.  This is because  their values are very small (see Sec.~\ref{sec:wtb}) compared to the size of the bounds obtained in our analysis  which are of the order of $10^{-2}-10^{-1}$. 

\section{Summary}\label{sec:3} 
In this work we have taken up the study of observables constructed out
of kinematical variables of top decay products for the purpose of
measuring top polarization in the presence of anomalous $Wtb$ couplings
as well as measuring the anomalous coupling $f_{2R}$ itself. We
concentrate on laboratory-frame variables which do not require the
reconstruction of the top rest frame. An
important consideration has been the degree of boost of the decaying
top, since for many practical processes, as for example, a heavy
resonance decaying into a top pair, the top quark is produced with
large momentum in the lab frame. 

We have considered four
observables -  asymmetries in the variables $\theta _{\ell},u,x_{\ell}$
and $z$. They are compared for their sensitivities to the polarisation
of the top quark and the anomalous coupling $f_{2R}$. We state the results 
of the comparison of asymmetries in two categories: 1. Asymmetries for 
the measurement the top-quark polarisation $P$, and 2. Asymmetries for 
the measurement of the anomalous coupling $f_{2R}$. 

As for the first category of asymmetries for 
the measurement of the top-quark polarisation, for small values 
of boost from the top quark rest frame to the lab frame 
($\beta\approx 0$), $A_{\theta _{\ell}}$ is the most sensitive observable.
Next in sensitivity is $A_z$ as long as $f_{2R}$ is small or 
negative. For large values of boosts ($\beta\sim 1$), $A_u$ and $A_z$ 
can be used as they are much more sensitive to $P$ compared to 
$A_{x_{\ell}}$ and $A_{\theta _{\ell}}$. 

For the second category corresponding to the measurement 
of the anomalous coupling $f_{2R}$, for all values of $\beta $, $A_z$
 can be used to measure $f_{2R}$ as long as $P\neq0$. 
For $P=0$, $A_u$, $A_z$ can be used to measure $f_{2R}$ 
for any $\beta$. The angular asymmetry $A_{\theta _{\ell}}$ is 
not suitable  as a measure of $f_{2R}$ for any value of $\beta$ 
as its sensitivity to $f_{2R}$ is much more smaller 
than the sensitivities of $A_{u}$ and $A_{z}$.  
Irrespective of the production mechanism of the top quark, $A_u$ 
can be used to measure $f_{2R}$ at large values of the boost $\beta$.  

In all cases, we determine the $1\sigma$ and $2\sigma$ limits that the measurement of asymmetries can put on the determination of the polarisation or $f_{2R}$ with a chosen number of events.  We also do an analysis of the use of combination of asymmetries for the
simultaneous determination of the top polarisation as well as $f_{2R}$. Finally we study the effects of including systematic uncertainties of asymmetries and find that for large values of top polarisation our observables  are sensitive to both $P$ and $f_{2R}$ for systematic uncertainties upto $\sim5\%$.

Note added: While our manuscript was in preparation a related work \cite{Cao:2015doa} appeared. In this work, correlations of the anomalous couplings $f_{1L,R}$ and $f_{2L,R}$  are obtained through global fits to data on observables that are insensitive to the polarisation of the top. They point out the need of measuring the single top cross section to $1\%$ precision as this would put strong constraints on the new physics that affects the $Wtb$ vertex. However, our method which uses observables sensitive to top polarisation, when used for processes such as single top production and decay, is sensitive to new physics even when the cross section is measured only to $5\%$. 
\appendix 

\section{The $x_{\ell}$ distribution} 

The
differential distribution $(1/\Gamma) d^2\Gamma/dx_{\ell,0}d\cos\theta
_{\ell,0}$ defined in the top quark rest frame is given by:
\begin{align} \label{eq:app0}
\frac{1}{\Gamma}\frac{d^2\Gamma}{dx_{\ell,0}dt_0}
&=\frac{3\xi ^2}{X}(1-x_{\ell,0})\big[f_{1L}^2\xi x_{\ell,0}(1+Pt_0)\\
\nonumber 
&+2f_{1L}\operatorname{Re}(f_{2R})\sqrt{\xi}(1+Pt_0)\\ \nonumber
&+|f_{2R}|^2\{Pt_0\left(\xi x_{\ell,0}+\frac{2}{x_{\ell,0}}-(\xi
+1)\right)\\ \nonumber 
&+(\xi +1)-x_{\ell,0}\xi\}\big] 
\end{align} 
where
$t_0=\cos\theta _{\ell,0}$ the cosine of the angle between the top spin
direction and the lepton momentum and $X$ is as defined in eq.
~\ref{eq:4}\footnote{We verified that upon integration over the azimuthal angles of the $b$-quark and the lepton all the structures of eq. A8 of \cite{Bernreuther199253,*Erratum} agree with those of our eq. ~\ref{eq:app0}. We have also checked that all the structures that appear in expression A8 of \cite{Bernreuther199253,*Erratum} are present in intermediate stages of the calculations that lead to eq.~\ref{eq:app0}.}. It is also the polar angle of the lepton due to our choice
of the top rest frame (see footnote in sec ~\ref{sec:int}). In the above equation, the top polarization points in the direction of motion of the top.  When the top polarisation points in a general direction in the top rest frame the differential distribution of the top decay is given, to linear order in $f_{2R}$ by:
\begin{align}\label{eq:app00}
\frac{1}{\Gamma}\frac{d\Gamma}{d x_{\ell,0} dt_0 d\phi _0 d\alpha _0}&=\frac{3\xi ^2}{X}\frac{1}{(2\pi)^2}(1-x_{\ell, 0})\\ \nonumber 
&\Big\{f_{1L}^2\xi x_{\ell,0}(1+\vec{P}\cdot\hat{p_{\ell,0}})\\ \notag
&+2\sqrt{\xi}\operatorname{Re}(f_{1L}f_{2R}^{\ast})\Big[\xi x_{\ell,0}(1+\vec{P}\cdot\hat{p}_{\ell,0})\\ \nonumber
&+(1-\xi x_{\ell})+\frac{1}{2}x_{\ell}(\xi -1)\vec{P}\cdot (\hat{p}_{b,0}-\hat{p}_{\ell,0})\Big]\\ \nonumber
&-\sqrt{\xi}\operatorname{Im}(f_{1L}f_{2R}^{\ast})x_{\ell,0}(\xi -1)\vec{P}\cdot(\hat{p}_{b,0}\times\hat{p}_{\ell,0})\Big\}
\end{align}
 where $\hat{p}_{\ell,0}$ and $\hat{p}_{b,0}$ are the unit vectors along the direction of momenta of the lepton and the $b$-quark in the top rest frame respectively; $\phi _{\ell,0}$ and $\alpha _0$ are the azimuthal angles of the lepton and the $b$-quark as mentioned in sec ~\ref{ssec:3}.

  Now we consider lab frame distributions. Let $\beta$
be the magnitude of the boost  required to go from the top rest frame to
the lab frame. The corresponding Lorentz transformation 
relates the energy and the polar angle of the lepton
measured in these two frames by
\begin{align} \label{eq:app1} 
x_{\ell} &=
\gamma (x_{\ell,0}+\beta x_{\ell,0}\cos\theta _{\ell,0})=\gamma
x_{\ell,0}(1+\beta t_0)\\ \nonumber x_{\ell}\cos\theta _{\ell} &= \gamma
(x_{\ell,0}\cos\theta _{\ell,0}+\beta x_{\ell,0}) = \gamma
x_{\ell,0}(t_0+\beta) 
\end{align} 
The inverse relations are
\begin{align} \label{eq:app2} 
x_{\ell,0} &= \gamma (x_{\ell}-\beta
x_{\ell}\cos\theta _{\ell})=\gamma x_{\ell}(1-\beta t)\\ \nonumber
x_{\ell,0}\cos\theta _{\ell,0} &= \gamma (x_{\ell}\cos\theta
_{\ell}-\beta x_{\ell}) = \gamma x_{\ell}(t-\beta) .
\end{align} 
Now the
differential distribution $(1/\Gamma) d^2\Gamma/dx_{\ell,0}dt_0$ is
transformed to $(1/\Gamma) d^2\Gamma/dx_{\ell}dt$ accordingly.
\begin{align}
\frac{1}{\Gamma}\frac{d^2\Gamma}{dx_{\ell}dt}&=\frac{3(1-\beta ^2)\xi
^2}{x_{\ell}(1-\beta t)^3X}(\gamma x_{\ell}(\beta t-1)+1)\\ \nonumber
&[f_{1L}^2\gamma ^2\xi x_{\ell}^2(\beta t -1)^2(P(t-\beta)-\beta t+1)\\
\nonumber 
&-2\gamma\sqrt{\xi}f_{1L}\operatorname{Re}(f_{2R})x_{\ell}(\beta
t-1)(P(t-\beta)-\beta t+1)\\ \nonumber 
&+|f_{2R}|^2(P(t-\beta)(\gamma
^2\xi x_{\ell}^2(\beta t-1)^2\\ \nonumber 
&+\gamma x_{\ell}(\xi
+1)(\beta t-1)+2)\\ \nonumber &+\gamma x(\beta t-1)^2(\xi(\gamma
x_{\ell}(\beta t-1)+1)+1))] 
\label{eq:app3} 
\end{align}

\begin{figure}
\includegraphics[keepaspectratio=true,scale=0.45]{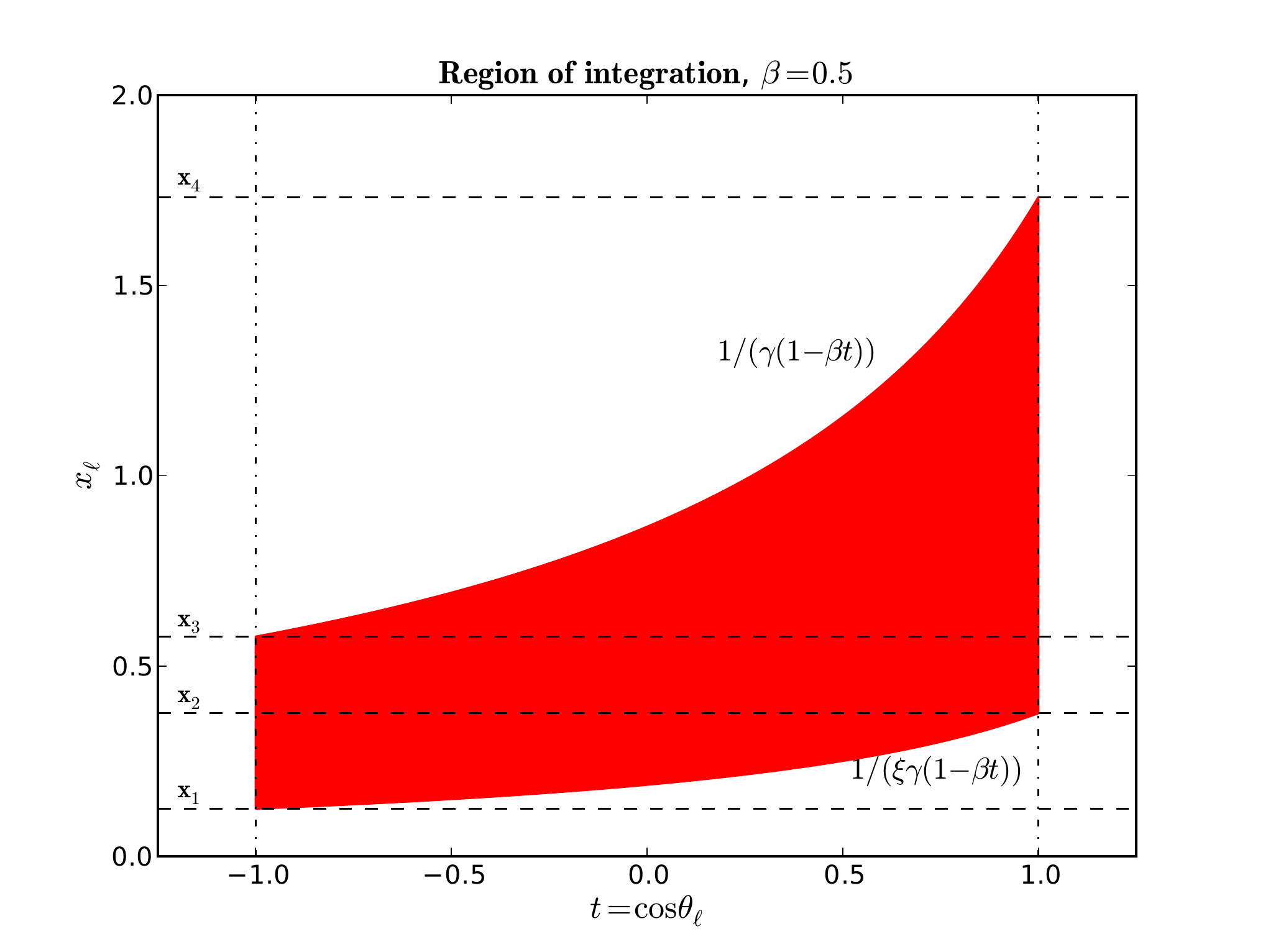}
\includegraphics[keepaspectratio=true,scale=0.45]{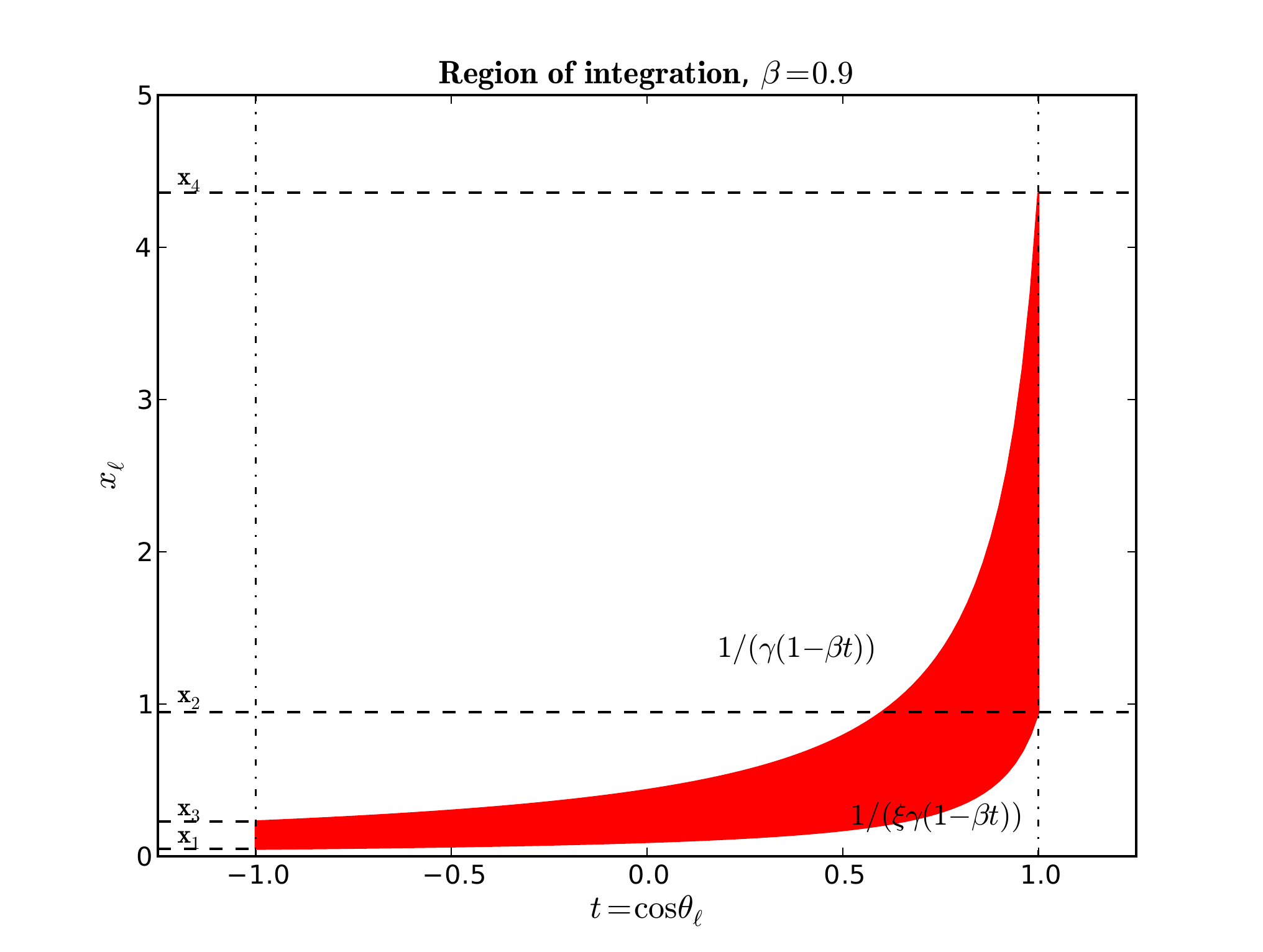}
\caption{Regions of integration for the $x_{\ell}$ distribution. The
left (right) figure corresponds to $\beta=0.5$ ($\beta=0.9$).}
\label{fg:app1} \end{figure} Integrating  the differential distribution
$(1/\Gamma) d^2\Gamma/dx_{\ell}dt$ over $t=\cos\theta _{\ell}$ in the
region bounded by eq.~\ref{eq:app2} gives the distribution
$(1/\Gamma) d\Gamma/dx_{\ell}$.

The region of integration is given in fig. ~\ref{fg:app1} for two
different value of the boost chosen such that the left (right) figure
corresponds to $\beta <\beta _c$ ($\beta>\beta _c$). $\beta _c=(\xi
-1)/(\xi +1)\approx 0.643$ is the value of the boost where the lowest
ordinate of the curve $x=1/(\gamma(1-\beta t))$ equals the maximum
ordinate of the curve $x=1/(\xi\gamma (1-\beta t))$. As shown in
fig. ~\ref{fg:app1} the range of $x_{\ell}$ is divided into three
regions, each having a separate integration limit on $t$. For $\beta
<\beta _c=(\xi -1)/(\xi +1)$, we have $[x_1,x_2],[x_2,x_3],[x_3,x_4]$
where, $x_1 =(1/\xi)\sqrt{(1-\beta)/(1+\beta)}$, $x_2
=(1/\xi)\sqrt{(1+\beta)/(1-\beta)}$, $x_3 =\sqrt{(1-\beta)/(1+\beta)}$,
$x_4 =\sqrt{(1+\beta)/(1-\beta)}$, For $\beta >\beta _c$, the range of
$x_{\ell}$ is divided into $[x_1,x_3],[x_3,x_2],[x_2,x_4]$ and for
$\beta=\beta _c$, $x_2=x_3$.  We first consider the case $\beta <\beta
_c$. The distribution $(1/\Gamma) d\Gamma/dx_{\ell}$ is called
$R1(x_{\ell})$ in the region $[x_1,x_2]$, $R2(x_{\ell})$ in the region
$[x_2,x_3]$ and $R3(x_{\ell})$ in the region $[x_3,x_4]$.
 
 The expressions are given below for ($f_{2R}=0$): \begin{align}
{}R1(x_{\ell}) &= -\frac{3\xi}{2\beta ^2(\xi -1)^2(\xi
+2)}\Big[(P-\beta)(1-2\xi)\sqrt{1-\beta ^2}\\ \nonumber &-2\xi P
x_{\ell}(1-\beta ^2)+x_{\ell}^{ 2}\xi ^2(P+\beta)\sqrt{1-\beta ^2}\\
\nonumber &+2\beta ^2 x_{\ell}^{ 2}\xi
^2(1-P)\sqrt{\frac{1+\beta}{1-\beta}}+2x_{\ell}\xi
^2(P-\beta)(1+\beta)\\ \nonumber &-x_{\ell}\xi ^2P(1-\beta
^2)\log\left(\frac{1+\beta}{1-\beta}\right)\\ \nonumber
&-2x_{\ell}\xi ^2 P(1-\beta
^2)\log(x_{\ell}\xi)\Big] \end{align}

\begin{align} R2(x_{\ell})&=\frac{6x_{\ell}\xi ^3}{(\xi +2)(\xi
-1)^2}[1- \frac{P}{\beta}-\frac{x_{\ell}}{\sqrt{1-\beta
^2}}+\frac{Px_{\ell}\beta}{\sqrt{1-\beta ^2}}\\ \nonumber
&+\frac{P(1-\beta ^2)}{\beta ^2}\tanh ^{-1}\beta] \end{align}

\begin{align} R3(x_{\ell})&=-\frac{3\xi ^3}{2\beta ^2(\xi -1)^2(\xi
+2)}\bigg[(P-\beta)\sqrt{1-\beta ^2} \\ \nonumber &-x_{\ell}^
2\left((P+\beta)\sqrt{1-\beta ^2}-2\beta
^2(1+P)\sqrt{\frac{1-\beta}{1+\beta}}\right)\\ \nonumber
&+2x_{\ell}(1-\beta)\Big(\beta(1+P)+P(1+\beta)\log x_{\ell} \\ \nonumber
&+\frac{1}{2}P(1+\beta)\log\left(\frac{1-\beta}{1+\beta}\right)\Big)\bigg]
\end{align} Similarly  for $\beta >\beta _c$, functions corresponding to
the intervals $[x_1,x_3]$,$[x_3,x_2]$,$[x_2,x_4]$ are called
$S1(x_{\ell}),S2(x_{\ell})$ and $S3(x_{\ell})$ respectively. We have
$S1(x _{\ell})=R1(x_{\ell})$ and $S3(x_{\ell})=R3(x_{\ell})$.

\begin{align} S2(x_{\ell})&= \frac{3\xi\sqrt{1-\beta ^2}}{2\beta ^2(\xi
-1)^2(\xi +2)}\Big[(\xi -1)^2(\beta -P)\\ \nonumber
&+2Px_{\ell}\sqrt{1-\beta ^2}(\xi ^2\log\xi -\xi (\xi -1))\Big]
\end{align}

\begin{acknowledgments}
RMG wishes to acknowledge support from the Department of Science and Technology, India under 
Grant No. SR/S2/JCB-64/2007. SDR acknowledges support from the Department of
Science and Technology, India, under the J.C. Bose National
Fellowship programme, Grant No. SR/SB/JCB-42/2009.

\end{acknowledgments}
\bibliography{ref} \bibliographystyle{apsrev4-1}

 \end{document}